\documentclass{article}
\usepackage[left=3cm, right=3cm, top=3cm, bottom=3cm]{geometry}
\usepackage[toc]{appendix}
\usepackage{amsmath}
\usepackage{graphicx}
\usepackage{hyperref}
\usepackage{color}
\usepackage{blindtext}
\usepackage{multicol}
\usepackage{authblk}
\usepackage{xcolor}
\usepackage{subfig}
\usepackage{cleveref}

\newcommand{\bs}[1]{\boldsymbol{#1}}
\newcommand{\pd}[2]{\frac{\partial #1}{\partial #2}}
\newcommand{\braket}[1]{\langle #1 \rangle}

\title{Large fluctuations in NSPT computations: \\
a lesson from $O(N)$ non-linear sigma models}
\author[a,b]{Paolo Baglioni}
\author[a,b]{Francesco Di Renzo}
\affil[a]{Dipartimento di Scienze Matematiche, Fisiche e Informatiche, Universit\`{a} di Parma}
\affil[b]{INFN, Gruppo Collegato di Parma}
\date{}

\begin{document}

\maketitle

\begin{abstract}
    In the last three decades, Numerical Stochastic Perturbation Theory (NSPT) has proven to be an excellent tool for calculating perturbative expansions in theories such as Lattice QCD, for which standard, diagrammatic perturbation theory is known to be cumbersome. Despite the significant success of this stochastic method and the improvements made in recent years, NSPT apparently cannot be successfully implemented in low-dimensional models due to the emergence of huge statistical fluctuations: as the perturbative order gets higher, the signal to noise ratio is simply not good enough. This does not come as a surprise, but on very general grounds, one would expect that the larger the number of degrees of freedom, the less severe the fluctuations will be. By simulating $2D$ $O(N)$ non-linear sigma models for different values of $N$, we show that indeed the fluctuations are tamed in the large $N$ limit, meeting our expectations: for a large number of {\em internal} degrees of freedom  ({\em i.e.} for large enough $N$), NSPT perturbative computation can be pushed to large perturbative orders $n$. By re-expressing our perturbative expansions as power series in the $gN$ ('t Hooft) coupling, we show some evidence that at any given order $n$ there is a tendency to gaussianity for the stochastic process distributions at large $N$. By summing our series, we can verify leading order results for the energy and its (field theoretic) variance in the large $N$ limit. We finally establish general relationships between the various perturbative orders in the expansion of the (field theoretic) variance of a given observable and combinations of variances and covariances of given orders NSPT stochastic processes.
    Having established all this, we conclude discussing interesting applications of NSPT computations in the context of theories similar to $O(N)$ ({\em i.e.} $CP(N-1)$ models).
\end{abstract}


\section{Introduction}

Lattice regularization of a field theory represents today one of the primary methods for investigating non-perturbative phenomena in theories that are too complex to study with analytical methods, such as QCD. The constant development of new MonteCarlo algorithms, combined with the improvements of hardware resources, has made non-perturbative numerical calculations accessible, which were previously beyond reach using non-numerical strategies. 
Quite interestingly, it is also possible to numerically implement stochastic algorithms for perturbation theory that are {\em in practice} quite close to the machinery of non-perturbative simulations. The most well-known and fruitful tool is Numerical Stochastic Perturbation Theory (NSPT). Initially introduced in the '90s \cite{DiRenzo1994} as a numerical implentation of Parisi and Wu's Stochastic Quantization \cite{Parisi1980}, today it is available in various (more or less) equivalent formulations \cite{DallaBrida2017, DallaBrida20172}. In particular, NSPT provides a direct pathway to the numerical computation of very general observables in Lattice QCD at perturbative orders that have never been explored analytically \cite{DiRenzo:2006qtj, Bali2013, DelDebbio2018}. NSPT’s success can (also) be attributed to its systematic and efficient numerical automation of order-by-order operations \cite{DiRenzo2004}, exactly in the same way as in what has become in recent times quite popular under the name of Automatic Differentiation (see {\em e.g.} \cite{Catumba2023}).

Since the initial introduction of NSPT, considerable expertise has been gathered, aiming in particular at a deeper comprehension of the underlying stochastic process. This interest was significantly triggered by the observation that, when applied to {\em small systems} (in a sense that will be clear in the following), NSPT produced huge fluctuations (not normally distributed) at orders that the method could successfully manage to compute for {\em larger ones} \cite{Alfieri2000}. 

\begin{figure}[h]
  \centering
  {\includegraphics[width=0.62\linewidth]{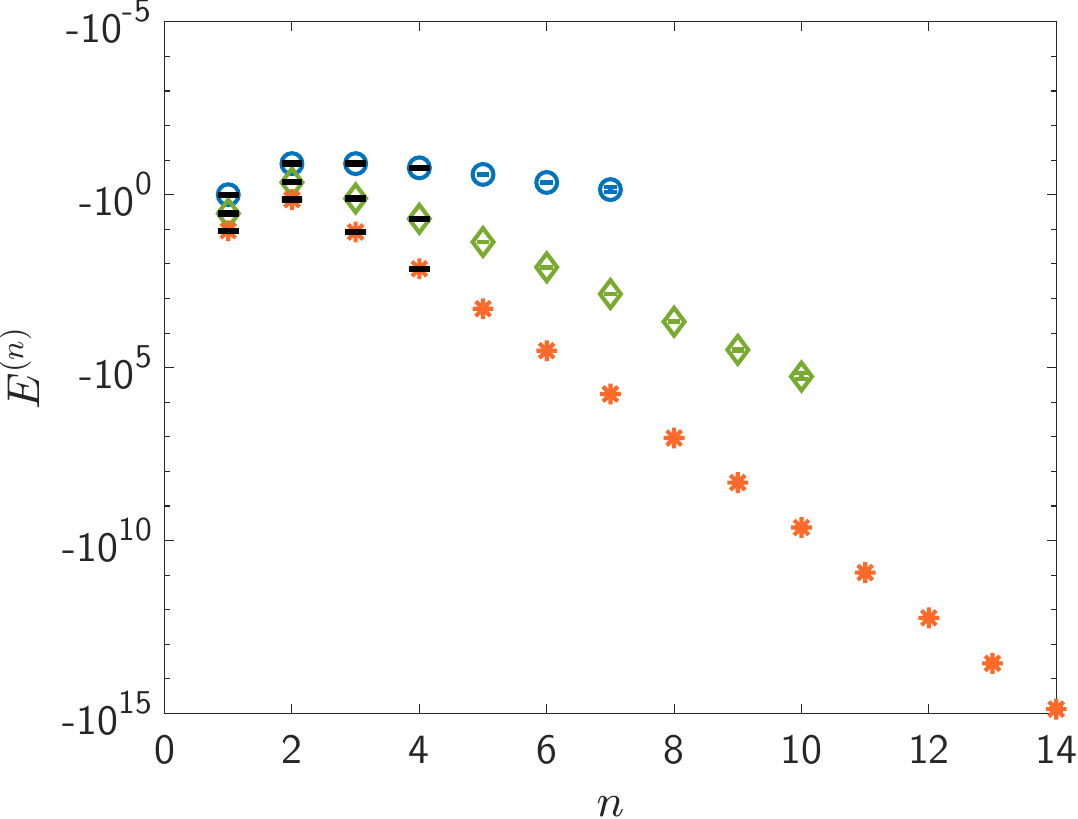}}
  \caption{The main message of this work as seen in NSPT perturbative computations in $O(N)$ non-linear sigma models: the energy of the model is computed at increasing perturbative order $n$;  blue symbols refer to $O(5)$, green ones to $O(15)$, orange ones to $O(45)$ (known analytic results are in black). While increasing fluctuations prevent us from safely computing high orders at {\em small} $N$, the larger is $N$, the higher loop we can safely compute.}
  \label{fig:MAINmessage}
\end{figure}

In this work we investigate how to make more quantitative the observation that the presence of large deviations at high perturbative orders can be related to the density of degrees of freedom of the system at hand. This conjecture finds an ideal testing ground in the study of  $O(N)$ non-linear sigma models (NLSM). A tendency of distributions of NSPT computed coefficients to Gaussian in a large $N$ limit has already been reported (at low orders) in \cite{GonzlezArroyo2019}. Our strategy will enable us to identify regimes ({\em i.e.} values of $N$) in which the model can be safely simulated and reliable predictions can be made. All in all, the scenario is that depicted in Fig. \ref{fig:MAINmessage}: NSPT computations are safe in the large $N$ limit (while at each value of $N$ we can afford reliable computations up to a maximum loop order $n$). A first account of this work has been presented in \cite{Baglioni:2024wyq}.

The results displayed in Fig. \ref{fig:MAINmessage} are not the only conclusions we reach. We will see that, switching to the $gN$ ('t Hooft) coupling, we can inspect a general tendency to gaussianity of the NSPT stochastic processes distributions as $N$ gets larger and larger. Expansions in the $gN$ coupling will be summed and shown to agree with results dictated by leading order large $N$ expansions. Finally, we will establish a general result for NSPT: for any given observable $O$, there are simple relationships between the various perturbative orders in the expansion of the (field theoretic) variance of $O$ and combinations of variances and covariances of (given orders of) the corresponding NSPT stochastic processes.\\

This work is structured as follows: in Section 1 we provide a brief description of NSPT and the motivations behind this study; the large deviations issue will be introduced and a simple, natural conjecture will be presented. In Section 2 we introduce two-dimensional $O(N)$ non-linear sigma models, revisiting in particular the setup for perturbation theory. In Section 3 we present the onset of fluctuations at not-so-large $N$ and how this effect is substantially tamed in the large $N$ limit. Section 4 provides a more quantitative description based on the study of relative errors for different values of $N$; by looking at the results of our study {\em a posteriori}, we will recognise criteria for assessing which $N$ is large enough for successful NSPT computations at a given loop order. In Section 5 we switch to the ('t Hooft) coupling $gN$, show how this enables to inspect a general (although slow) tendency of 
NSPT stochastic processes to gaussianity for large $N$ and compare the results we obtain by summing our perturbative expansions with leading order large $N$ results. In Section 6 we obtain general results relating variances and covariances of NSPT stochastic processes to perturbative expansions of field theoretic variances. Finally, in Section 7 we conclude with some remarks and future prospects, in particular outlining potential opportunities for the study of asymptotic effects ({\em e.g.} renormalons, expansions around non-trivial vacua, resurgence).

\subsection{Numerical Stochastic Perturbation Theory}

NSPT was initially inspired by the famous work of Parisi and Wu \cite{Parisi1980} on Stochastic Quantization (SQ). With this respect, NSPT is basically a numerical implementation of stochastic perturbation theory (which was originally approached in a diagrammatic form; for a review see \cite{Damgaard1987}). In short, NSPT is nothing more than the numerical integration of a tower of Langevin equations rewritten order-by-order, {\em i.e.} in perturbation theory. The mechanism underlying NSPT is actually more general and can be implemented starting from many stochastic differential equations \cite{DallaBrida2017,DallaBrida20172}. In the following, we will adhere to the original formulation.\\
To be more precise, we can sketch the path from SQ to NSPT as follows. We are interested in calculating expectation values of a Euclidean field theory, namely
\begin{align}
    \label{eq:euclidean_average}
    \langle O[\phi]\rangle = \frac{\int \mathcal{D}\phi\ O(\phi)\ e^{-\mathcal{S}_E[\phi]}}{\int \mathcal{D}\phi\ e^{-\mathcal{S}_E[\phi]} }
\end{align}
The basic idea is to promote the fields to functions of a new extra variable (known in the literature as stochastic time)
\begin{align}
    \phi(\bs{x}) \rightarrow \phi_{\eta}(\bs{x},\tau)
\end{align}
 and evolve the system in this fictitious time according to the Langevin stochastic equation
 \begin{align}
    \label{eq:langevin}
     \frac{d \phi_{\eta}(\bs{x},\tau)}{d\tau} = -\frac{\partial\mathcal{S}_E[\phi]}{\partial\phi_{\eta}(\bs{x},\tau)} + \eta(\bs{x},\tau)
 \end{align}
 where $\eta(\bs{x},\tau)$ is a normalized Gaussian noise satisfying
 \begin{align}
     \langle \eta(\bs{x},\tau) \rangle_\eta = 0 \quad  \langle \eta(\bs{x},\tau)\eta(\bs{x}',\tau') \rangle_\eta = 2\delta(\bs{x}-\bs{x}')\delta(\tau-\tau')
 \end{align}
In the previous equations, the symbol $\langle \dots \rangle_{\eta}$ means 
\begin{align}
    \label{eq:average_over_noise}
    \langle \dots \rangle_\eta = \frac{\int \mathcal{D}\eta \ \dots \ e^{-\frac{1}{4} \int d\bs{x} d\tau \eta^2(\bs{x},\tau)} }{\int \mathcal{D}\eta \ e^{-\frac{1}{4} \int d\bs{x} d\tau \eta^2(\bs{x},\tau)}}
\end{align}
while the subscript in $\phi_{\eta}(\bs{x},\tau)$ denotes the dependence on the realization of the noise. 
The fundamental assertion of SQ is that the expectation value of an observable with respect to the noise in Eq. \eqref{eq:average_over_noise} reproduces, in the limit of infinite stochastic time, the expectation values of the Euclidean field theory we are interested in Eq.~\eqref{eq:euclidean_average}, that is
\begin{align}
    \lim_{\tau\to\infty} \langle O[\phi_\eta(\bs{x}_1,\tau) , \phi_\eta(\bs{x}_2,\tau), \dots, \phi_\eta(\bs{x}_n,\tau)]\rangle_\eta = \langle O[\phi(\bs{x}_1), \phi(\bs{x}_2), \dots, \phi(\bs{x}_n)]\rangle
    \label{eq:convergence}
\end{align}
Because of Eq. \eqref{eq:convergence}, the numerical integration of Eq. \eqref{eq:langevin} provides one of the algorithmic options for non-perturbative Monte Carlo simulations \cite{Ukawa1985, Batrouni1985}. In this context, the systematic error arising from the introduction of a finite time step can be eliminated by an additional \textit{à la} Metropolis acceptance step \cite{Roberts1998}. On the other hand, NSPT takes the solution of Eq. \eqref{eq:langevin} and expand it in a (formal) power series in the coupling constant (from now on, we will omit the subscript $\eta$ for convenience):
\begin{align}
\label{eq:pt_exp_field}
    \phi(\bs{x},\tau) = \phi^{(0)}(\bs{x},\tau) + \sum_{n=1}^{\infty}g^n \phi^{(n)}(\bs{x},\tau)
\end{align}
It is easy to see that, by substituting the previous expression into Eq.~\eqref{eq:langevin} and recollecting the various orders in the coupling constant $g$, we obtain a tower of partial differential equations
\begin{align}
\label{eq:order_by_order_langevin}
\begin{split}
     \frac{d \phi^{(0)}(\bs{x},\tau)}{d\tau} & = -G_0^{-1}\phi^{(0)}(\bs{x},\tau) + \eta(\bs{x},\tau) \\
     \frac{d \phi^{(1)}(\bs{x},\tau)}{d\tau} & = -G_0^{-1}\phi^{(1)}(\bs{x},\tau) + D_1(\phi^{(0)}) \\
     \dots\\
      \frac{d \phi^{(n)}(\bs{x},\tau)}{d\tau} & = -G_0^{-1}\phi^{(n)}(\bs{x},\tau) + D_n(\phi^{(0)},\phi^{(1)}, \dots, \phi^{(n-1)})
\end{split}
\end{align}
where $G_0^{-1}$ is the free propagator ({\em i.e.}, $\frac{\partial\mathcal{S}_0[\phi]}{\partial\phi^{(0)}(\bs{x},\tau)} = G_0^{-1}\phi^{(0)}(\bs{x},\tau)$, where $S_0$ is the free Euclidean action) and $D_n$ the interaction terms that mix different perturbative orders of the fields.  It is worth noting that the truncation is exact at any  order: the equation at a given perturbative order $n$ only depends on the fields at order $n' \leq n$. Of course, the tower of equations must be integrated numerically with a given integrator and a given value of a time step. With this respect it is important to point out a practical difference between the non-perturbative solution of the Langevin equation and NSPT: in the latter case, we cannot introduce an acceptance step to eliminate systematic errors in the time step size. Instead, we will extrapolate results in the $\Delta\tau\to 0$ limit, the appropriate type of extrapolation depending on the order of the integrator used. In recent times the non exact character of the perturbative numerical solution of Eq. \eqref{eq:order_by_order_langevin} has been debated. All in all, this is inherent in the nature of the perturbative expansion mechanism: the latter makes sense only provided an analytic solution exists, which is unavoidably lost in a process like an accept/reject mechanism. Obviously, many integration schemes can be adopted, whose relative merits have been discussed \cite{DallaBrida2017,DallaBrida20172}. In this work we considered the simplest possible choice, namely the Euler integrator. For further details and motivations the reader is referred to App. \ref{app:tau_estrap}. 

\subsection{Large deviations in NSPT simulations}
\label{subsec:pepe_eff}

Since the early times of NSPT it was clear that the ability to estimate perturbative coefficients using a stochastic method unavoidably exposes the results to statistical and systematic errors. We have already referred to errors coming from the numerical integration of the stochastic equation. Since we unavoidably have to work in finite volume, one should also care of finite volume effects. Certainly significant progress has been made with respect to both these \cite{Brambilla:2013sua, Bali2013, Bali2014, DallaBrida2017, DallaBrida20172}. On the other hand, the distributions of NSPT coefficients themselves reveal several features and peculiarities that must be handled with great care. This observation was triggered by some discrepancies between NSPT results and known ones for $O(3)$ non-linear sigma model \footnote{This observation was made by M. Pepe \cite{Pepe1996} and that's why the effect is often referred to as {\em Pepe effect}.} and was subsequently investigated and characterized in detail for zero-dimensional models (in particular for the zero-dimensional $\lambda\phi^4$ model, the dipole random variable model and the Weingarten’s ``pathological model'') \cite{Alfieri2000}. The scenario that emerges is that, for all of them, the statistical properties of the process deviate significantly from those of a normal process, which exhibits exponentially suppressed tails. In contrast, NSPT stochastic processes at higher orders are characterized by long tails and rare events that introduce {\em spikes} that are difficult to handle with conventional statistical analysis. Despite the fact that estimating statistical errors can be done more safely adopting non-parametric methods (in particular the bootstrap method \cite{Alfieri2000}), one is nevertheless left with the practical challenge of understanding if (and to which extent) one can get precise estimates at high perturbative orders for low-dimensional systems. \\
The possible appearance of large fluctuations at high orders is not surprising; looking at a simple model, even a simple-minded argument can provide a hint. Let us consider the zero-dimensional action:
\begin{align}
    S = \frac{1}{2}\varphi^2 + \frac{1}{3}g \varphi^3
\end{align}
with the associated Langevin equation
\begin{align}
    \dot{\varphi} = -(\varphi + g \varphi^2) + \eta
\end{align}
Considering the perturbative expansion in Eq. \eqref{eq:pt_exp_field} we obtain the following tower of equations:
\begin{align}
\begin{split}
\label{eq:phi3_pt_eq}
    \dot{\varphi}^{(0)} & = - \varphi^{(0)} + \eta \\
    \dot{\varphi}^{(1)} & = - \varphi^{(1)} - \varphi^{(0)}\varphi^{(0)} \\
    \dot{\varphi}^{(2)} & = - \varphi^{(2)} - 2\varphi^{(0)}\varphi^{(1)} \\
    \dot{\varphi}^{(3)} & = - \varphi^{(3)} - (2\varphi^{(0)}\varphi^{(2)} + \varphi^{(1)}\varphi^{(1)}) \\
    \dots
\end{split}
\end{align}
It is clear that any oscillation dictated by the Gaussian noise at the leading order propagates squared at the first order, results in a cubic effect at the third order and so on. All in all, a potentially large fluctuation is magnified at higher and higher perturbative orders, and we are naturally led to consider the possibility that this mechanism can ultimately make the signal a wild one. Will the restoring mechanism which is built into the Langevin equation be effective enough to reabsorb large fluctuations? How long will this mechanism take and what are the effects on the autocorrelation time of the process? Are we guaranteed that a finite standard deviation will emerge? These are examples of the questions that we naturally ask ourselves. A natural hypothesis is of course that this mechanism will be less and less severe when more degrees of freedom come into play (coherent large fluctuations will be unlikely and all in all the coupling of the many degrees of freedom are expected to eventually result in some - maybe slow - convergence to gaussianity). Of course, then again the question arises of how large a system should be to stay on a safe side. 

This well-known problem has recently come back in the framework of NSPT computation of perturbative expansions around non-trivial vacua. This formulation of NSPT has to be considered a natural one (in the same spirit, there is work on the Schr{\"o}dinger Functional \cite{Torrero2010, Hesse2014}) and would be very useful, given the intricate nature of standard perturbation theory in this context. Some preliminary steps have been taken, starting from low-dimensional models like the Double Well Potential) \cite{Baglioni2023}. Not surprisingly, fluctuations were back; actually they are the biggest hurdles in assessing the feasibility of these calculations.

\subsection{A testing ground for a natural conjecture}
\label{sec:conj}

The structure shown in Eq. \eqref{eq:phi3_pt_eq} is quite general, but the fact that we can expect less problems for a large number of degrees of freedom is apparently confirmed by the case of Lattice QCD, where fluctuations do eventually take place at very high orders, but these are beyond those needed, for example, to inspect asymptotic behaviors.
In large systems the order-by-order equations are coupled and, along the same simple lines as those pinned down in section \ref{subsec:pepe_eff}, one expects that rare events affecting a single degree of freedom independently of the others will contribute less. All in all, it is natural to investigate the relations between NSPT stochastic distributions and the number of degrees of freedom. From this perspective, simulating $O(N)$ non-linear sigma models is a natural choice, as we can consider the same model for different values of the parameter $N$, changing the number of degrees of freedom. Certainly, having more and more degrees of freedom translates into a larger computational cost, so we have to prove we can get positive answers before too high computational costs are to be paid.
\section{ \texorpdfstring{$O(N)$}{} non-linear sigma models}

$O(N)$ non-linear sigma models provide a nice theoretical laboratory in quantum field theory. From a theoretical point of view, NLSM display properties that are crucial in understanding the dynamics of particles and fields, such as asymptotic freedom. From a phenomenological point of view, they have been able to model different features in different contexts (for an introduction to the model see {\em e.g} \cite{zinn-justin2002}). In the following we will not been concerned with any particular physical motivation, our interest for the model being motivated by the possibility to tune the parameter $N$, changing the number of degrees of freedom. In the continuum, we have the following action:
\begin{align}
    S = \frac{1}{2g}\int d\bs{x}\ \Bigl(\partial_\mu \bs{s}(\bs{x}) \Bigr)^2
\end{align}
where $\bs{s}(\bs{x})$ is a $N$-component real scalar field constrained by $\bs{s}(\bs{x})\cdot \bs{s}(\bs{x}) = 1$ for all $\bs{x}$ (from now on, we will denote the point in space-time without bold). On the lattice, we can provide various discrete versions of this theory. In this work we use the simplest $2D$ formulation, namely
\begin{align}
    S = -\frac{1}{g}\sum_{x,\mu} \bs{s}_x \cdot \bs{s}_{x+\mu} \quad \bs{s}_x \cdot \bs{s}_x = 1
\end{align}
where $\bs{s}_x$ is a $N$-component lattice real scalar field constrained by $\bs{s}_x\cdot \bs{s}_x = 1$ for all the lattice sites, $g$
is the coupling constant and $\mu$ runs over the two lattice directions. The partition function of the system can be written by incorporating the constraint into a local Dirac delta function
\begin{align}
\label{eq:PF}
    Z = \int \prod_{x}\ d\bs{s}_x \ \delta(\bs{s}^2_x-1) \ e^{\frac{1}{g}\sum_{x,\mu} \bs{s}_x \cdot \bs{s}_{x+\mu}}
\end{align}

\subsection{Perturbation theory}

Perturbation theory requires correctly identifying the degrees of freedom. Following the standard approach \cite{Elitzur1983}, one way consists of eliminating the constraints by decomposing
\begin{align}
    \bs{s}_x = (\bs{\pi}_x,\sigma_x)
\end{align}
and rescaling
\begin{align}
    \bs{\pi}^2_x \rightarrow g\bs{\pi}^2_x
\end{align}
This way, we can integrate out the constraint in the partition function in Eq. \eqref{eq:PF}, remaining only with the degrees of freedom associated with $\bs{\pi}_x$, which are now unconstrained. Neglecting insignificant terms in perturbation theory, we can write the partition function 
\begin{align}
\label{eq:PF_PT}
    Z = \lim_{\lambda\to 0}\int \prod_{x}\ d\bs{\pi}_x \ e^{-\frac{1}{2}\sum_{x,\mu} \Bigl[ (\Delta_\mu\bs{\pi}_x)^2 + \lambda^2 \bs{\pi}_x^2 -\frac{1}{g}(\Delta_\mu\sqrt{1-g\bs{\pi}_x^2})^2 \Bigr] - \frac{1}{2}\sum_x \log{(1-g\bs{\pi}_x^2)}}
\end{align}
where $\Delta_\mu$ is the usual lattice discretized version of the derivative; the logarithmic term arises from the additional integral measure term and $\lambda$ is an infrared regulator, which needs to be removed at the end of the calculations. This latter term is required because the propagator is ill-defined in the infrared region, diverging as $\log{\lambda}$ (see \cite{Elitzur1983}). 

Perturbation theory, as usual, amounts to expanding the interaction terms in Eq. \eqref{eq:PF_PT} in Taylor series and evaluating them on the free theory. This results in a highly intricate perturbation theory, where not only new Feynman diagrams are generated but also new vertices appear at each order \cite{Alles1997}. 
In this work we consider a well-defined $O(N)$ invariant quantity (\cite{Elitzur1983}), namely the energy (which is nothing but than the propagator in term of the original fields)
\begin{align}
\label{eq:energy}
    E & = -\frac{1}{2V} \pd{\log{Z}}{\Bigl(\frac{1}{g}\Bigr)}= \braket{ \boldsymbol{s}_0\cdot \boldsymbol{s}_1} = g \braket{ \boldsymbol{\pi}_0\cdot \boldsymbol{\pi}_1} + \braket{\sqrt{1 + g\boldsymbol{\pi}^2_0}\sqrt{1 + g\boldsymbol{\pi}^2_1}}
\end{align}
(where a perturbative evaluation requires the Taylor series expansion of the square root).

\subsection{NSPT setup for \texorpdfstring{$O(N)$}{} NLSM}

Calculating Eq. \eqref{eq:energy} in perturbation theory is a formidable task and up to now only the first four terms are known analitically \cite{Elitzur1983, Alles1997}. In the perturbative expression
\begin{align}
    E & = E^{(0)} + g E^{(1)} + g^2 E^{(2)} + g^3 E^{(3)} + g^4 E^{(4)} + \dots 
\end{align}
the loop corrections read
\begin{align}
\begin{split}
\label{eq:know_res}
    E^{(0)} & = 1 \\
    E^{(1)} & = -(N-1)/4 \\
    E^{(2)} & = -(N-1)/32 \\
    E^{(3)} & = -0.00726994(N-1)-0.00599298(N-1)^2 \\
    E^{(4)} & = -0.00291780(N-1)-0.00332878(N-1)^2 -0.00156728 (N-1)^3 \\
\end{split}
\end{align}
With NSPT we can proceed straight ahead and go beyond the fourth order, especially (as already stated and as we will see) in the large $N$ regime. In fact, NSPT simulations are completely insensitive to the increasing number of terms of the diagrammatic perturbation theory: the order-by-order encoding of Eq. \eqref{eq:order_by_order_langevin} is naturally generated. \\

\noindent Eq.~\eqref{eq:PF_PT} can be rewritten as
\begin{align}
\begin{split}
\label{eq:z_lambda_0}
    Z = \int \prod_{x}\ d\boldsymbol{\pi}_x\ & \exp{ \Bigl\{ \sum_{x,\mu} \Bigl( \boldsymbol{\pi}_x \cdot \boldsymbol{\pi}_{x+\mu} + \frac{1}{g} \sqrt{1-g\boldsymbol{\pi}_x^2}\sqrt{1-g\boldsymbol{\pi}_{x+\mu}^2} \Bigr)\Bigl\}} \\
    &\times\exp{ \Bigl\{ -\frac{1}{2}\sum_x \log(1-g\bs{\pi}_x^2 )\Bigr\} } 
\end{split}
\end{align}
From the action
\begin{align}
\label{eq:action}
    S = -\sum_{x,\mu} \Bigl( \boldsymbol{\pi}_x \cdot \boldsymbol{\pi}_{x+\mu} + \frac{1}{g} \sqrt{1-g\boldsymbol{\pi}_x^2}\sqrt{1-g\boldsymbol{\pi}_{x+\mu}^2} \Bigr) +\frac{1}{2}\sum_x \log(1-g\bs{\pi}_x^2 )
\end{align}
it is straightforward to derive the associated Langevin equation 
\begin{align}
\label{eq:langevin_continuum}
\begin{split}
    \dot{\pi}_y^j(\tau) = \sum_{\mu} \Biggl\{ \pi^j_{y+\mu} + \pi^j_{y-\mu} - \pi^j_y\Biggl( \sqrt{\frac{1-g\bs{\pi}_{y+\mu}^2}{1-g\bs{\pi}_{y}^2}} + \sqrt{\frac{1-g\bs{\pi}_{y-\mu}^2}{1-g\bs{\pi}_{y}^2}} \Biggr)\Biggr\}\Bigg|_{\bs{\pi}(\tau)}
    + \frac{g\pi^j_y}{1-g\bs{\pi}_y^2}\Bigg|_{\bs{\pi}(\tau)} + \eta_y^j(\tau)
\end{split}
\end{align}
$\pi_y^j(\tau)$ represents the $j$-component of the $\bs{\pi}$ field at the lattice site $y$, evaluated at the stochastic time $\tau$. In the Euler scheme, the equation reads
\begin{align}
\label{eq:langevin_euler}
\begin{split}
    \pi_y^j(\tau + \Delta\tau) = \pi_y^j&(\tau) + \Delta\tau\sum_{\mu} \Biggl\{ \pi^j_{y+\mu}(\tau) + \pi^j_{y-\mu}(\tau) \\
    &- \pi^j_y\Biggl( \sqrt{\frac{1-g\bs{\pi}_{y+\mu}^2}{1-g\bs{\pi}_{y}^2}} + \sqrt{\frac{1-g\bs{\pi}_{y-\mu}^2}{1-g\bs{\pi}_{y}^2}} \Biggr)\Biggr\}\Bigg|_{\bs{\pi}(\tau)} + \Delta\tau\frac{g\pi^j_y}{1-g\bs{\pi}_y^2}\Bigg|_{\bs{\pi}(\tau)} + \sqrt{2\Delta\tau} \eta_y^j(\tau)
\end{split}
\end{align}
The Gaussian white noise is distributed according to a normal distribution with zero mean and unit variance (and now no Dirac delta is around). Again, after Eq.~\eqref{eq:pt_exp_field} has come into play at a given fixed-order, Eq. \eqref{eq:langevin_euler} gets translated into a tower of (perturbative) equations.

In Eq. \eqref{eq:z_lambda_0}, \eqref{eq:action}, \eqref{eq:langevin_continuum} and \eqref{eq:langevin_euler} we consider directly the limit $\lambda\to 0$ so that the stochastic evolution exhibits a zero-mode that must be regularized. We have opted for the simplest scheme, {\em i.e.} a step-by-step subtraction of the zero-mode (see App. \ref{app:zero_mode} for a discussion).
\section{Fluctuations: how they show up and how they are tamed in the large \texorpdfstring{$N$}{} limit}
\label{sec:fluctuations}
\begin{table}[t]
    \centering
    \begin{tabular}{|lllll|}

    \multicolumn{1}{|c|}{$\bs{N}$} & \multicolumn{1}{c|}{$\bs{n_{max}}$} & \multicolumn{1}{c|}{$\bs{V}$} & \multicolumn{1}{c|}{\textbf{Statistics $\times N$}} & \multicolumn{1}{c|}{$\bs{\Delta\tau}$} \\
    \hline
    \multicolumn{1}{|c|}{$5:1:15$}  & \multicolumn{1}{c|}{$15$} & \multicolumn{1}{c|}{$20\times20$} & \multicolumn{1}{c|}{$\approx 1.6 \cdot 10^9 $} & $[18,\ 25,\ 35,\ 50,\ 75,\ 100] \cdot 10^{-4
}$\\
    \multicolumn{1}{|c|}{$18:3:45$}  & \multicolumn{1}{c|}{$15$} & \multicolumn{1}{c|}{$20\times20$} & \multicolumn{1}{c|}{$\approx 2.1 \cdot 10^9 $} & $[18,\ 25,\ 35,\ 50,\ 75,\ 100] \cdot 10^{-4
}$\\
    \multicolumn{1}{|c|}{$15:3:43$} & \multicolumn{1}{c|}{$23$} & \multicolumn{1}{c|}{$20\times20$} & \multicolumn{1}{c|}{$\approx 1.2 \cdot 10^9 $} & $[18,\ 25,\ 35,\ 50,\ 75,\ 100] \cdot 10^{-4
}$\\
    \multicolumn{1}{|c|}{$45$}      & \multicolumn{1}{c|}{$23$} & \multicolumn{1}{c|}{$20\times20$} & \multicolumn{1}{c|}{$\approx 2.7 \cdot 10^{9} $} & $[18,\ 25,\ 35,\ 50,\ 75,\ 100] \cdot 10^{-4
}$\\
    \multicolumn{1}{|c|}{$5$}      & \multicolumn{1}{c|}{$23$} & \multicolumn{1}{c|}{$66\times66$} & \multicolumn{1}{c|}{$\approx 9 \cdot 10^7 $} & $[18,\ 25,\ 35,\ 50,\ 75,\ 100] \cdot 10^{-4
}$\\

    \end{tabular}
    \caption{Simulation details. The notation $N_1:m:N_2$ stands for $N_1$, $N_1+m$, $N_1+2m$, \dots, $N_2$. 6 different time step sizes $\Delta\tau$ were considered. $n_{max}$ indicates the highest perturbative order in a given set of simulations; notice that streams with different $n_{max}$ have to be considered separately because of correlations among the different orders in a given stream. Statistics is normalized by $N$, {\em i.e.} different rows have approximately the same computational weight.}
    \label{tab:simulation_details}
\end{table}

\begin{figure}[htb]
  \centering
  {\includegraphics[width=.5\linewidth]{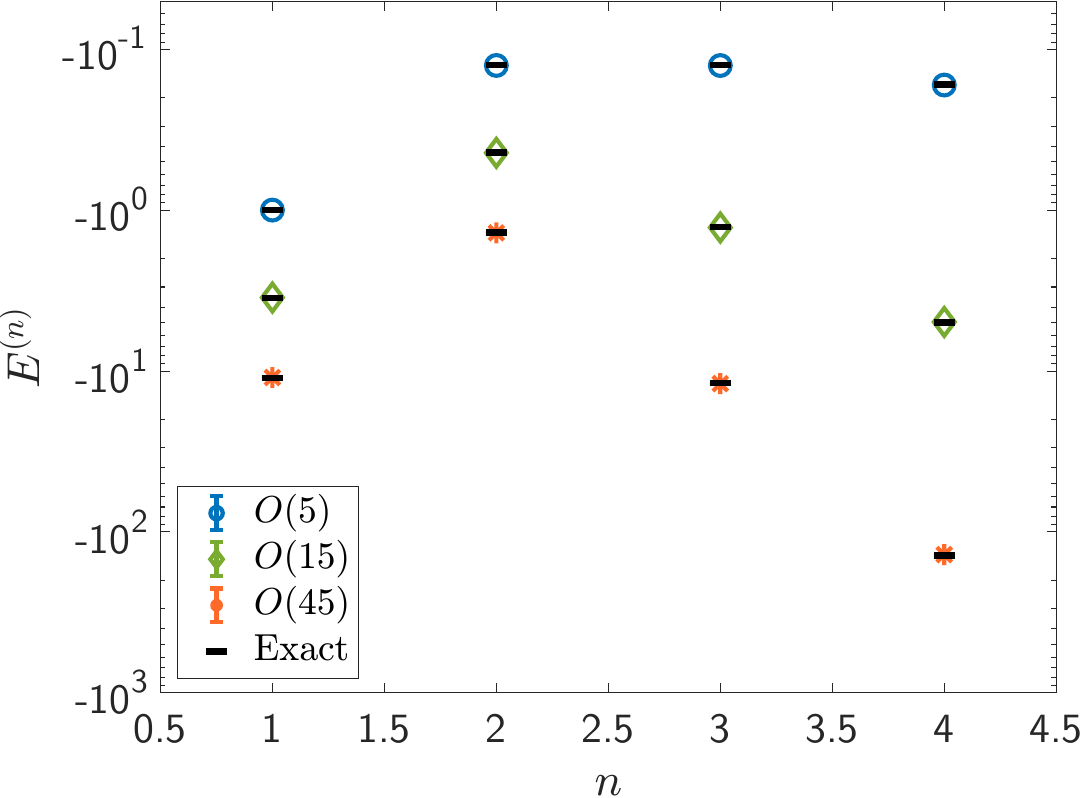}}
  \caption{NSPT computations of the energy for $O(5)$ (blue markers), $O(15)$ (green markes)  and $O(45)$ (orange markers) up to the fourth order; analytical values are plotted in black. The errorbars are not visible as they are smaller than the markers. Computations were performed on a $20\times20$ lattice.}
  \label{fig:know_val_comp}
\end{figure}

We performed different simulations for several values of $N$ on a $20\times20$ lattice, unless otherwise specified (see Tab. \ref{tab:simulation_details} for further details). Even on such a small lattice size we found tiny finite volume effects as far as the first four known values are concerned (see Eq. \eqref{eq:know_res} \cite{Alles1997}): discrepancies are of the order of a few per mille. In Fig. \ref{fig:know_val_comp} we plot NSPT predictions and exact results. Systematic effects coming from the choice of the numerical integration scheme in Eq. \eqref{eq:langevin_euler} have been removed. Details about extrapolations (which asks for dealing with the minimization of a $\chi^2$ and the estimation of autocorrelations) are discussed in App.~\ref{app:tau_estrap} and App. \ref{app:autocorr}. 

Having looked at Fig.~\ref{fig:know_val_comp}, the reader is invited to look back at Fig. \ref{fig:MAINmessage}, where one can easily grasp the effect of $N$ on the maximum perturbative order we could compute. The difference between the low and the high perturbative order ($n$) regions is evident: for all the values of $N \geq 5$ which are taken into account, we could compute three extra orders on top of the ones that were already known, but as $n$ increases, we were able to obtain reliable results only for increasing $N$. More precisely, for the $O(5)$ NLSM model we pushed the calculation from the fourth up to the seventh order, for $O(15)$ we reached the tenth order, and for $O(45)$ even the 14th order was calculated. As already said, this is related to the fluctuations in the stochastic process, in a way that will be evident in the figures that we are going to display in the following. It is not trivial to determine at which perturbative order $n$ one will have to stop for a given $O(N)$ NLSM model. In the following we will provide at least a few sanity checks by looking at time series, cumulative moving averages and standard deviations and (later on) scaling of (estimations of) relative errors. 

\begin{figure}[t]
  \centering
  {\includegraphics[width=1.0\linewidth]{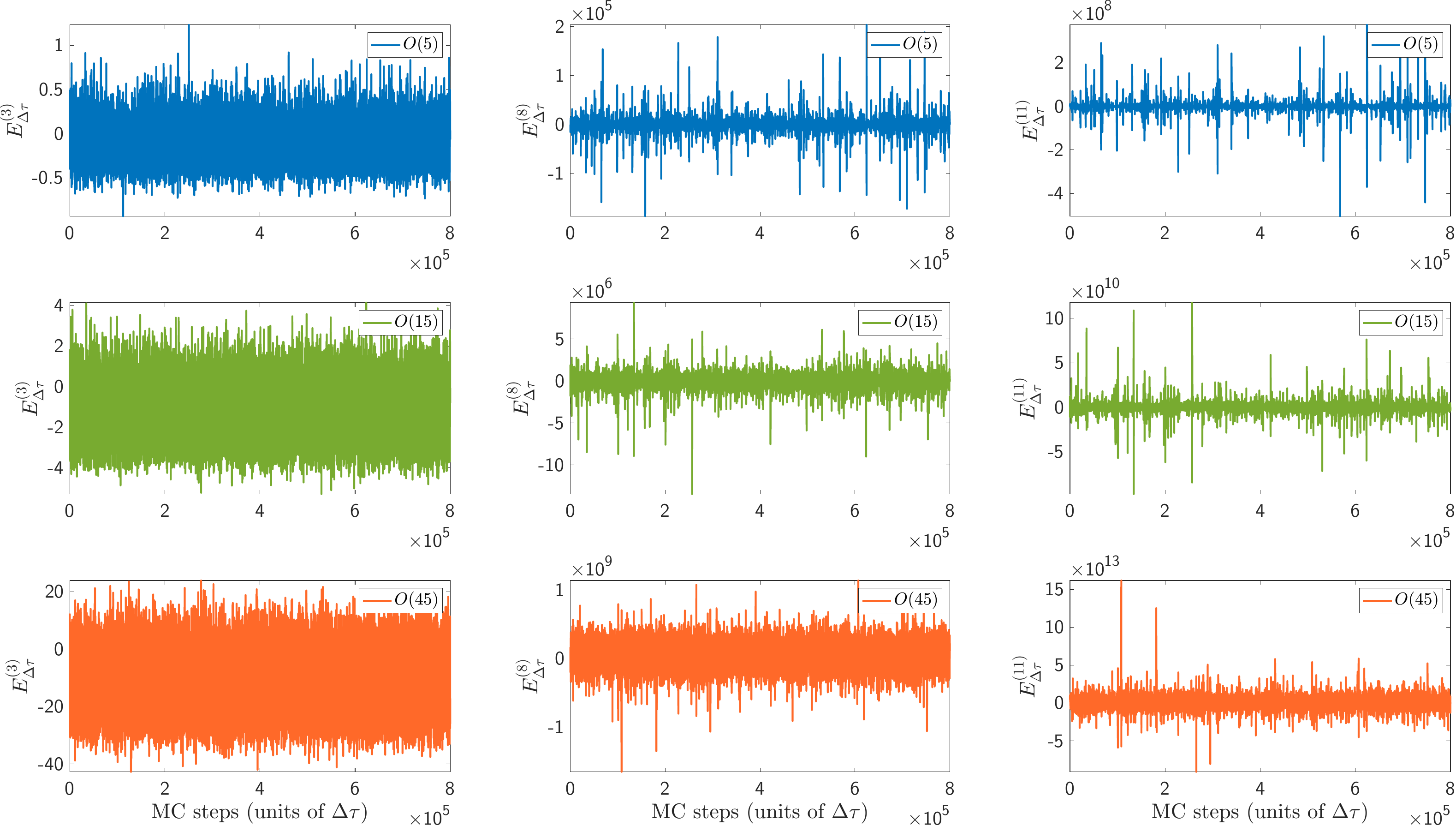}}
  \caption{NSPT signals for the $O(5)$ (blue lines), $O(15)$ (green lines) and $O(45)$ (orange lines) models at $\Delta\tau = 0.0035$ for $n=3$ (left), $n=8$ (center) and $n=11$ (right). We considered for all models  $800000$ steps, after having previously subtracted $4000$ thermalization steps (a number chosen looking at the highest order).}
  \label{fig:signal}
\end{figure}

The fact that large fluctuations show up is evident by inspection of Fig.~\ref{fig:signal}, where we display the signals ({\em i.e.} stochastic time series) of NSPT simulations for $O(5)$, $O(15)$ and $O(45)$ (same color code as in Fig.~\ref{fig:MAINmessage}): comparison is made at the same value of $\Delta\tau = 0.0035$. In the first column, we show the evolution in stochastic time of the signals at $n=3$: no sign of large fluctuations is there. By studying the distribution of the signals, one finds that these are (very) similar to a Gaussian process, not exhibiting long tails. Most importantly, no significant differences are there for different values of $N$. This is no longer true at perturbative order $n=8$ (second column): large spikes emerge for the $O(5)$ model which contribute significantly to the mean and standard deviation of the distribution (in section \ref{sec:cum} we report a more complete characterization). On the other hand, the signal for $O(15)$ still seems manageable, and that for $N=45$ looks excellent. Notice that most of the fluctuations (at any value of $n$ and $N$) are distributed within what looks like a {\em band}: if we want to inspect the presence of large fluctuations by eye, we have to compare them to the width of this band. An even more extreme situation shows up in the last column (perturbative order $n=11$). For $N=5,15$ we inspect large fluctuations, which result in significant deviations from the gaussian distribution and in a signal which appears bad in terms of signal to noise ratio. On the other hand, the stochastic time history for $N=45$ (last row) appears under control (remember: we have to compare the size of fluctuations to the width of the band in which most of them fall); we fell comfortable with the distributions.

\subsection{Cumulative moving averages and standard deviations}
\label{sec:cum}
\begin{figure}[t]
  \centering
  {\includegraphics[width=1.0\linewidth]{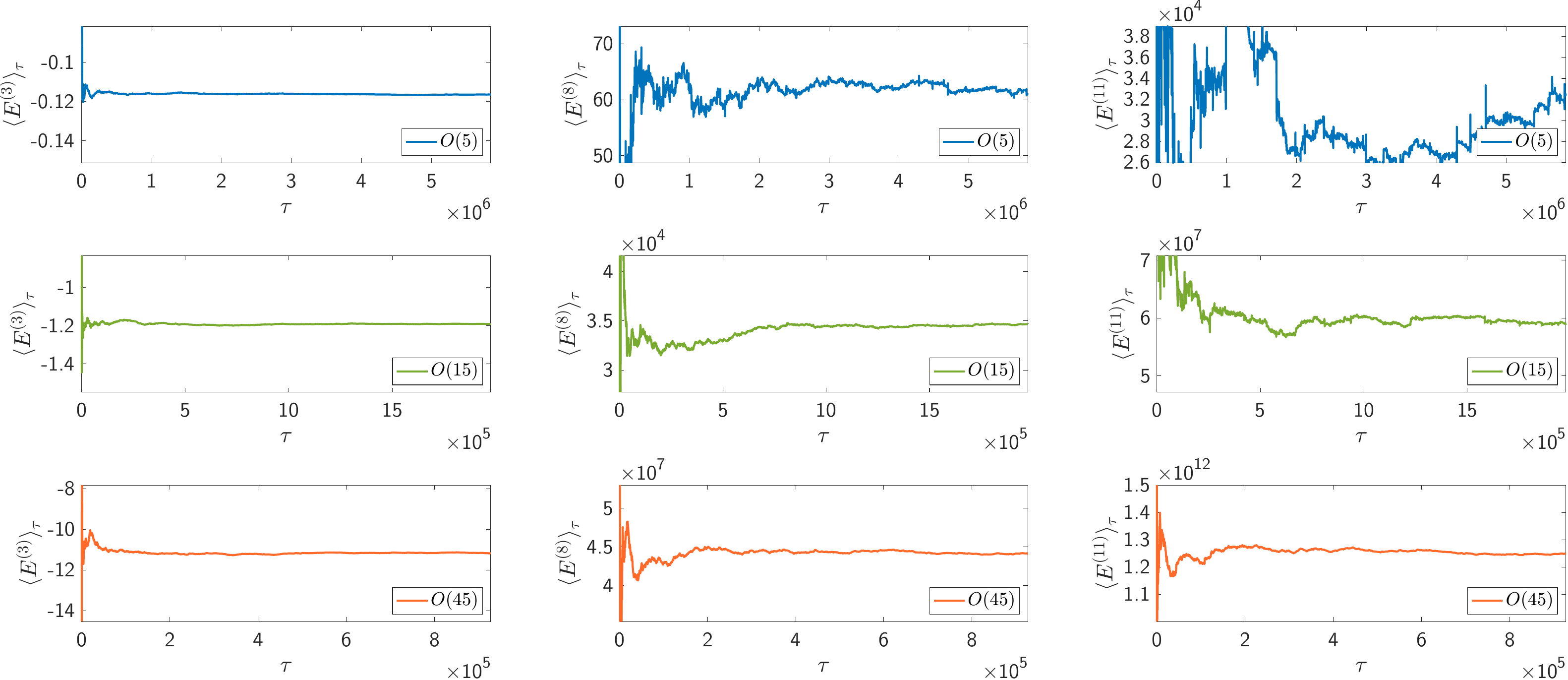}}
  \caption{Time evolution of the mean (cumulative moving averages) for $O(5)$ (blue lines), $O(15)$ (green lines) and $O(45)$ (orange lines). In all cases, $\Delta\tau = 0.01$. Comparisons are made at equal computational time, i.e. with $N_{stat} \cdot N_{dof} \sim \mathrm{const}$ ($N_{stat}$ being the number of time steps). $y$-value ranges have been chosen as $[\hat{y}-\delta \hat{y},\hat{y}+\delta \hat{y}]$, where $\hat{y}$ is the best estimation of the quantity at hand and $\delta \hat{y}$ is a spread amounting to a given percentage variation (which is the same, at each perturbative order, for all the signals that we compare).}
  \label{fig:cumsum}
\end{figure}

\begin{figure}[t]
  \centering
  {\includegraphics[width=1.0\linewidth]{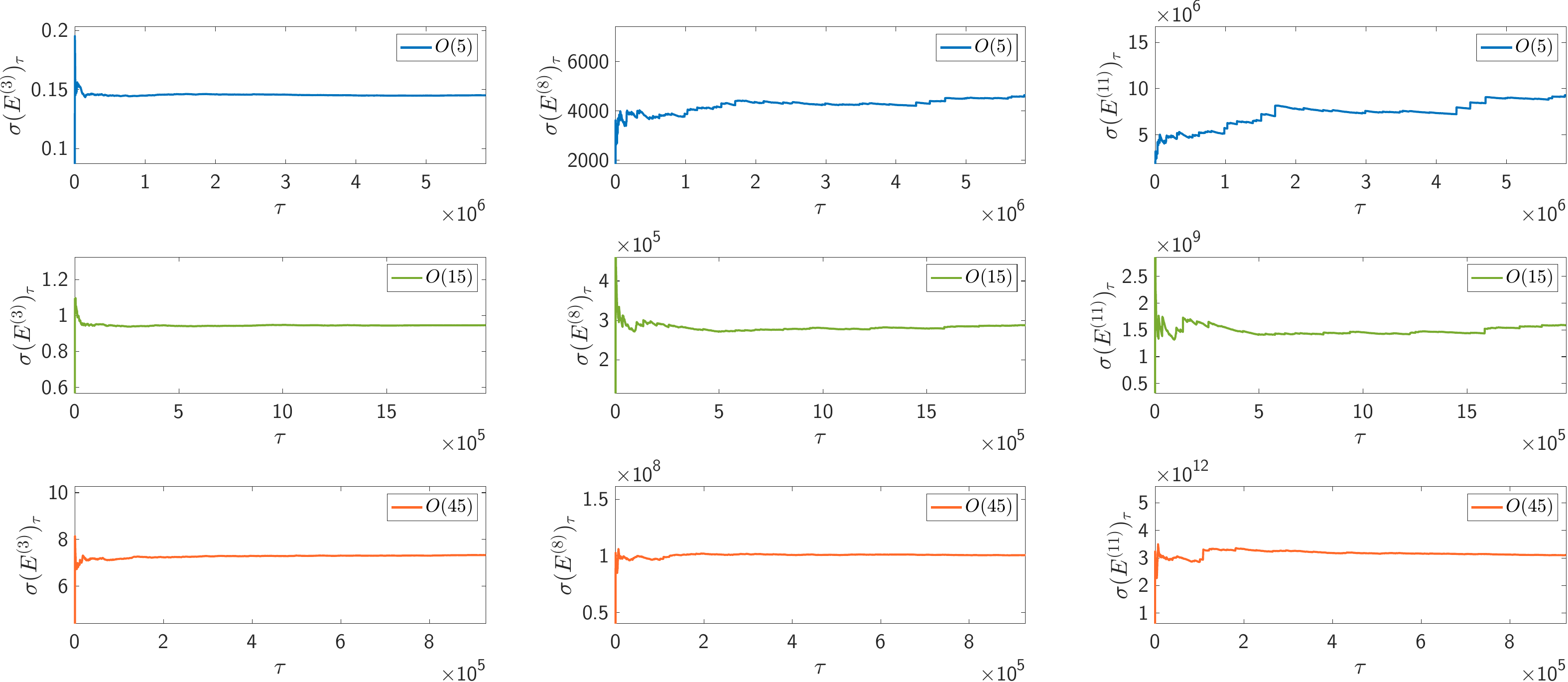}}
  \caption{Time evolution of the standard deviation (cumulative moving standard deviation) for $O(5)$, $O(15)$ and $O(45)$; $\Delta\tau = 0.01$. Same plotting conventions as in Fig.~\ref{fig:cumsum}.}
  \label{fig:cumstd}
\end{figure}

Cumulative moving average and standard deviation are statistical tools that can be helpful in analysing our data: we take them into account in the same spirit as in \cite{Alfieri2000}. Roughly speaking, the cumulative (moving) counterparts of mean and standard deviation evolve as more data points are added and track how data spread out as new observations are considered. This is a most natural way of assessing data stability with respect to the occurrence of (large) fluctuations. We define the cumulative moving average (in short, cumulative mean) as
\begin{align}
\label{eq:cum_mean}
    \braket{E^{(n)}}_\tau = \frac{1}{\tau}\sum_{t=1}^{\tau}E^{(n)}_t
\end{align}
where $E^{(n)}_t$ stands for the $n$-th perturbative order of the energy measured on the $t$-th Monte Carlo configuration (as it is easily seen, we make use of the same wording as for standard, non-perturbative Monte Carlo simulations; keep in mind that here a configuration is a collection of values of the field at different orders, at a given stochastic time). The subscript $\braket{...}_\tau$ indicates that the averaging window extends from the first configuration to the $\tau$-th. Similarly, we can define the cumulative standard deviation
\begin{align}
    \sigma(E^{(n)})_\tau = \sqrt{\braket{E^{(n)^2}}_\tau - \braket{E^{(n)}}_\tau^2}
    \label{eq:cum_std}
\end{align}
A good Monte Carlo simulation, accurately exploring a well-defined distribution, should exhibit for $\tau\gg 1$ an approximately constant (asymptotic) value for the quantities in Eq. \eqref{eq:cum_mean} and \eqref{eq:cum_std}. To make a fair comparison between results obtained at various values of $N$, which are expected to span different orders of magnitude (see Fig.~\ref{fig:MAINmessage}), it is a good idea to focus on relative fluctuations with respect to the estimated mean. This motivated our choice of how to plot Fig.~\ref{fig:cumsum} and Fig.~\ref{fig:cumstd}: $y$-value ranges have been chosen as $[\hat{y}-\delta \hat{y},\hat{y}+\delta \hat{y}]$, where $\hat{y}$ is the best estimation of the quantity at hand and $\delta \hat{y}$ is the spread amounting to a given percentage variation (which is the same, at a given order, for all the signals that we compare).  Additionally, larger values of $N$ are computationally more expensive to simulate but are expected to result in better self-averaging effects, so that we decided to compare different Monte Carlo histories at the same computational time rather than at the same statistics. The cumulative means and standard deviations that we display have been obtained by time histories at the same value of $\Delta\tau = 0.01$ (the effect is roughly the same at different values of $\Delta\tau$; this should appear clear by looking back at Fig.~\ref{fig:signal}, where the same overall picture emerged from comparisons made with the different choice of $\Delta\tau = 0.0035$).

The scenario that emerges for the quantity defined in Eq. \eqref{eq:cum_mean} is consistent with what was broadly seen before, as it can be seen from  Fig.~\ref{fig:cumsum}. The latter is organised exactly as Fig.~\ref{fig:signal}: on a given column, the perturbative order is fixed (as before, $n=3$, $n=8$ , $n=11$), while $N$ stays the same on a given row (as before, $N=5$, $N=15$, $N=45$). At a low perturbative order (first column, $n=3$), at any value of $N$ the cumulative means are seen to flatten, as we would expect them to do. On the contrary, the picture changes dramatically as the loop order increases. In the middle column ($n=8$) the cumulative mean for $O(5)$ does not flatten that well; it is subject to significant fluctuations and even with a substantial amount of statistics we are not sure we are getting a reliable estimate of the mean). It does not take too long for the signal for $O(15)$ to display a similar effect: see the last column ($n=11$). For $O(45)$ the cumulative mean can be considered stable and, since this observation is independent of $\Delta\tau$, a subsequent $\Delta\tau\to 0$ extrapolation turns out to be well under control. 

For smaller values of $N$ the scenario is even worse when analyzing the standard deviation (see Fig.~\ref{fig:cumstd}, which is plotted with the same conventions as in Fig.~\ref{fig:cumsum}). Obviously, due to the definition of the standard deviation, any fluctuation gives only positive contributions. All in all, for $N=5,n\ge8$ and $N=15,n\ge11$ we cannot be sure that our estimations of the variance are reliable. However, in this case as well, we can inspect well flattening signals for large enough $N$. Actually, for $O(45)$ we could not detect pathological effects up to $n=14$. As another sanity check, we verified that for $N=45$ the statistical errors are observed to scale correctly as $\sim \frac{1}{\sqrt{N_{samples}}}$. \\
All in all, we have clear hints that for given values of $N$ our best estimations of mean and standard deviation could be unreliable. We will see that these findings are confirmed and a more clear description appears by looking at the scaling of (best estimations of) relative errors: we present this analysis in section~\ref{sec:rel_err}.

\subsection{Large volumes and large \texorpdfstring{$N$}{}}

A key concept for the reliability of a lattice Monte Carlo simulation is the property of lattice self-averaging: for a given theory, as the size of the lattice grows, one has to observe more stable statistical averages.  We expect that using larger and larger lattices for computations of local quantities (as in Eq. \eqref{eq:energy}) results in a reduction of the standard deviation. On the other hand, we have just seen that, at a given NSPT order, fluctuations are tamed with increasing $N$ ({\em i.e.} with increasing number of local degrees of freedom). Naively, one could ask to which extent the two effects are related: not surprisingly, we will see that these two effects are fundamentally different.

\begin{figure}[hb]
  \centering
  {\includegraphics[width=1.0\linewidth]{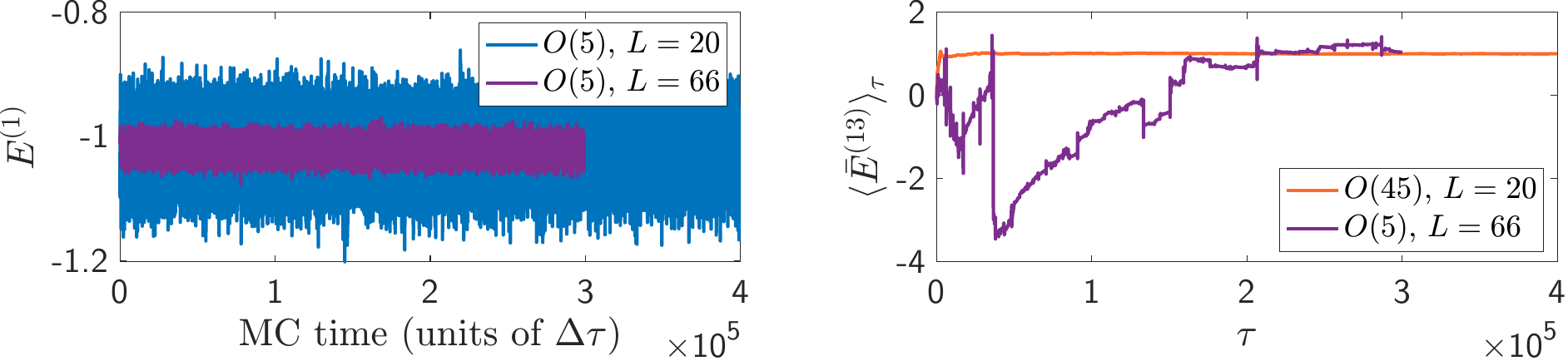}}
  \caption{Left plot: $O(5)$ model at the lowest, $n=1$ perturbative order for different lattice sizes ($L^2 = 66 \times 66$ and $L^2 = 20 \times 20$). Right plot: a comparison between the cumulative averages of the the $n=13$ perturbative order for $O(5)$ on a $66\times 66$ lattice and $O(45)$ on a $20\times 20$ lattice. The two models have (roughly) the same total number of degrees of freedom.}
  \label{fig:l66}
\end{figure}

In this regard, we can inquire about the \textit{true} size of the system (and inspect the effects on what we could broadly think of as NSPT self-averaging properties). Naively, one could think of considering the same total number of degrees of freedom (namely, we can compare the large $N$ - small $L$ regime with the small $N$ - large $L$ regime). In the left plot of Fig.~\ref{fig:l66} we inspect the expected self-averaging effect for $O(5)$. Namely, we compare stochastic time histories (stochastic time series) at $n=1$ order on $L^2 = 66 \times 66$ and $L^2 = 20 \times 20$: as expected, considering a larger lattice results in less statistical fluctuations. In the right plot of Fig.~\ref{fig:l66} we consider a much higher order, namely $n=13$. Here we compare NSPT results for $O(5)$ on a lattice volume $L^2 = 66 \times 66$ with NSPT results for $O(45)$ on $L^2 = 20 \times 20$: in this comparison the overall number of degrees of freedom is (roughly) the same. Given the polynomial dependence in $N$ (see Eq. \eqref{eq:know_res} at order $n=1,2,3,4$), we expect the two signals (for $O(5)$ and $O(45)$) to differ significantly in orders of magnitude: to compare them we need to normalize the data. We decided to look at cumulative means normalized to our best estimates. In practice, at each stochastic time we divide the cumulative means by the mean value calculated over the entire sample (we denote with a bar the quantities normalized in this way: $\braket{\Bar{E}^{(n)}}_\tau$). It is evident that the large $L$ - small $N$ combination does not help that much for computing high perturbative orders: taming large deviations appears to be a genuine large $N$ effect.
\section{Probing NSPT fluctuations with relative errors}
\label{sec:rel_err}
In Monte Carlo simulations, studying relative errors scaling is very useful for assessing the robustness of the results; of course NSPT is no exception. In this section, we make use of very general assumptions about the behavior of relative errors as functions of the two parameters $N$ and $n$. Through this analysis, we show how to identify {\em a posteriori} $N$ regions where NSPT results appear to be reliable and regions where the accuracy is lost: all in all, we want once again to assess the meaning of \textit{large} $N$ for a NSPT computation of a $O(N)$ quantity at a given perturbative order $n$.

\subsection{\texorpdfstring{$n$}{}-scaling}

From now on, we denote relative errors as
\begin{align}
\label{eq:rel_err}
    \Delta^{(n)}_N = \frac{\delta E^{(n)}}{E^{(n)}}\bigg|_{N}
\end{align}
For a given $O(N)$ we study the ratio of the value of the energy at the $n$-th perturbative order $E^{(n)}$ and the associated error $\delta E^{(n)}$. Notice that both are coming from the stochastic time extrapolation procedure that we describe in App. B and C. In particular, they are supposed to contain all the information regarding auto-correlations and cross-correlations. We stress that we compute the ratio in Eq.~\eqref{eq:rel_err} (and later on in Eq.~\eqref{eq:rel_err_bar}) out of our {\em best estimations} of $E^{(n)}$ and $\delta E^{(n)}$. Since we have already seen that problems are around at high orders, we must be ready to accept that we did our best, but some of these estimations will be unreliable; as a matter of fact, this is what will happen. First of all, we will analyze the quantity in Eq.~\eqref{eq:rel_err} as a function of the integer parameter $n$ for fixed values of $N$. 

A natural general hypothesis is that relative errors should be monotonically increasing in $n$: all in all, we have already seen that higher-order computations result in progressively larger standard deviations as $n$ increases. Part of this will also come from non-trivial effects of cross-correlations as a function of $n$ (a given order of the field depends on all the fields at lower order as indicated in Eq.~\eqref{eq:order_by_order_langevin}).  

\begin{figure}[htb]
  \centering
  {\includegraphics[width=1.0\linewidth]{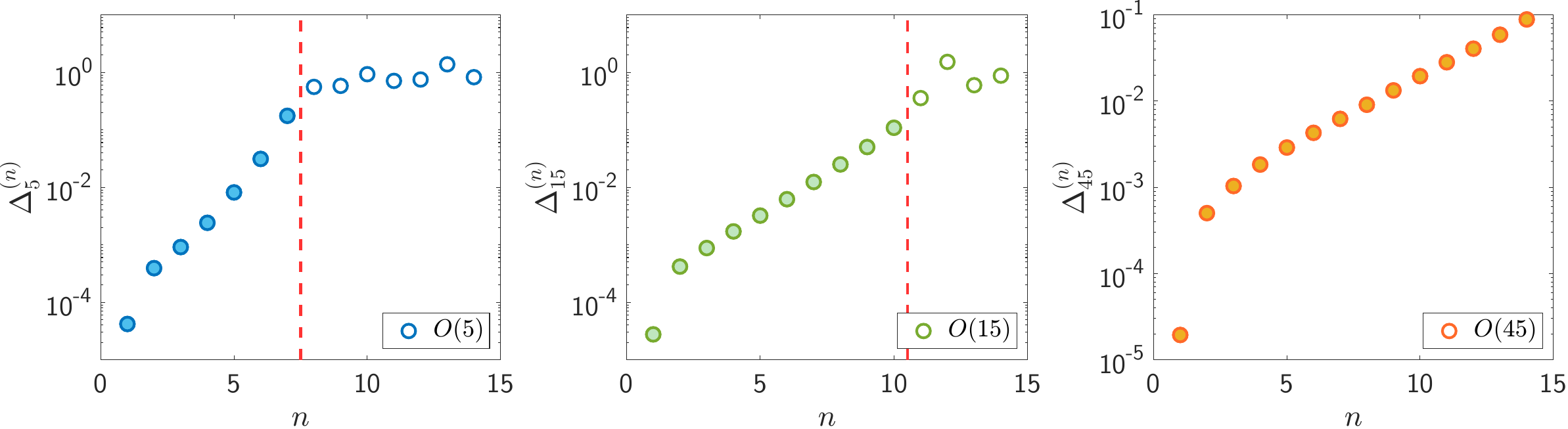}}
  \caption{Evaluation of the relative error in Eq. \ref{eq:rel_err} as a function of loop order for $O(5)$, $O(15)$ and $O(45)$. The solid markers indicate the results that we could regard as safe: the red dashed line separates safe and unsafe regions. Thresholds are the same as those guessed from the analysis of cumulative means and standard deviations. }
  \label{fig:n_small}
\end{figure}

In Fig.~\ref{fig:n_small} we plot $\Delta^{(n)}_N$ as a function of $n$ for various values of 
$N$: the three panels once again refer to $O(5)$, $O(15)$ and $O(45)$. As far as one sees a smooth behavior, one inspects an exponential growth. Notice that up to $n=14$ for $O(45)$ we observe no violations of our hypothesis: relative errors are monotonically increasing. The same does not hold for lower values of $N$, in particular for $n > 7$ in the $O(5)$ case and for $n>10$ in the $O(15)$ case. These thresholds seem to be in agreement with what has been discussed in section \ref{sec:cum} and are yet another reason for plotting Fig. \ref{fig:signal}. We have already made the point that we have been working with {\em best estimations} and we ended up with having to give up with some of them. As an extra piece of information, notice that we had also found less smooth continuum stochastic time extrapolations for the perturbative orders which turned out to be unreliable.

\subsection{\texorpdfstring{$N$}{}-scaling}

\begin{figure}[hbt]
  \centering
  {\includegraphics[width=1.\linewidth]{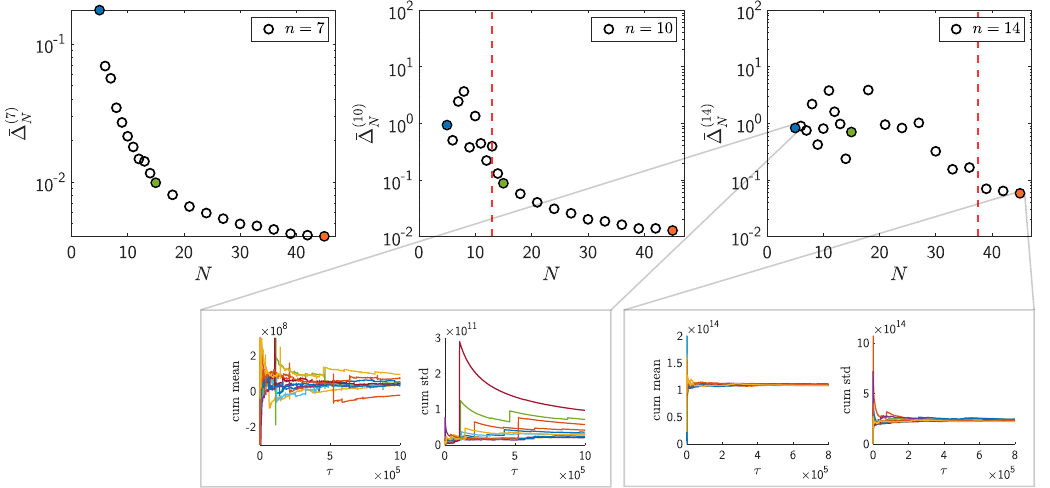}}
  \caption{Scaling of the relative errors defined in Eq. \eqref{eq:rel_err_bar} as a function of the integer parameter $N$ for all the models (from $O(5)$ to $O(45)$) listed in Tab.~\ref{tab:simulation_details}; the three plots refer to perturbative orders $n=7, 10, 14$. The filled points refer to $O(5)$, $O(15)$, and $O(45)$ (same color code as before, in particular as in Fig.~\ref{fig:MAINmessage}). The red dashed line separates the regions where we inspect the expected scaling from the regions where we see violations. To make contact with what has already been shown, in the small panels we show cumulative means and standard deviations for $10$ different independent runs for $O(5)$ and $O(45)$ at the $14$-th perturbative order.}
  \label{fig:n_big}
\end{figure}

It appears natural to look also for an $N$-scaling analysis of relative errors. Consider the quantity 
\begin{equation}
\label{eq:rel_err_bar}
    \bar{\Delta}^{(n)}_N = \frac{\delta E^{(n)}}{E^{(n)}}\bigg|_{N}\cdot \Gamma(N)    
\end{equation}
The definition is the same as before, apart for the presence of the factor $\Gamma(N)$. The latter is in charge of correcting for different statistics. In other words, we compare relative errors for different $N$ and the same amount of independent measurements, and thus the only difference in statistics which is left is that connected to $N$ itself.

On very general grounds we claim that the quantity in Eq. \eqref{eq:rel_err_bar} must exhibit a monotonically decreasing behavior in $N$ (all in all, after taking the factor $\Gamma(N)$ into account, we are left with different numbers of degrees of freedom depending on $N$). This is perfectly observed at small values of $n$, in particular for $n\leq 7$: see for $n=7$ the left plot of Fig.~\ref{fig:n_big} ($O(5)$, $O(15)$ and $O(45)$ results are highlighted with the same color code we have been using till now). For $n=10$ (central panel), the expected monotonic behavior only sets in for $N>12$, while for $n=14$ we need to look at the very largest values of $N$ to inspect the expected behavior. \\

All in all, we have been looking at different indicators in order to inspect what value of $N$ is large enough for a given perturbative order $n$. We want to stress that, whatever indicator we considered, we have always found the same answer. To further stress this, the reader is invited to have a look at the right panel ($n=14$) of Fig.~\ref{fig:n_big}. $O(5)$ and $O(45)$ results sit in different regions: only $O(45)$ results are candidate as reliable. Now look at the small panels. To make contact to what we have already looked at, we plot cumulative means and standard deviations for the energy at perturbative order $n=14$ for $O(5)$ and $O(45)$ as obtained from several independent simulations: the difference is impressive.

\section{Expansions in 't Hooft coupling: deeper into Large \texorpdfstring{$N$}{} analysis}

\subsection{Switching to 't Hooft coupling}

\begin{figure}[ht]
  \centering
  {\includegraphics[width=0.62\linewidth]{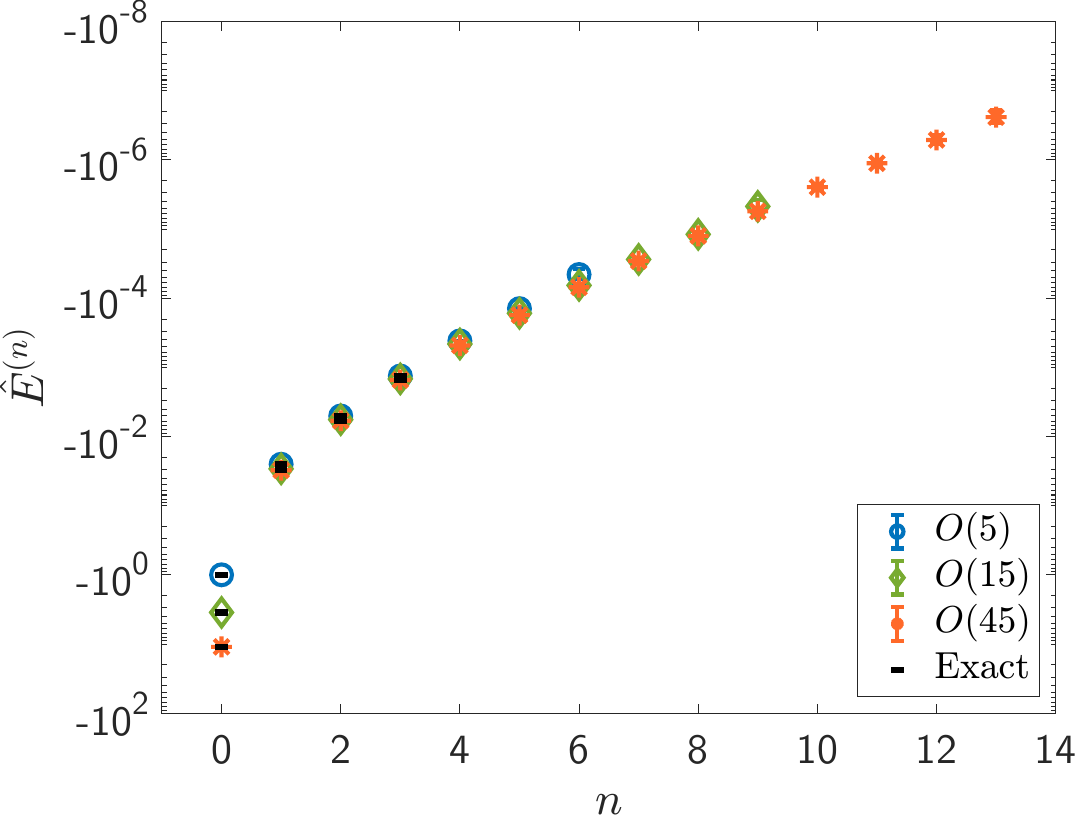}}
  \caption{The observable $\hat{E}$ (see Eq. \eqref{eq:hatE}) at increasing perturbative order $n$; same color code as in  Fig. \ref{fig:MAINmessage}: blue symbols refer to $O(5)$, green ones to $O(15)$, orange ones to $O(45)$ (known analytic results are in black).}
  \label{fig:MAINmessageNEW}
\end{figure}

Having provided enough evidence of a much better behaviour (much smallest fluctuations) of our perturbative computations at large $N$, we take a step forward with the introduction of a new coupling, à la 't Hooft. We now consider the observable
\begin{equation*}
\hat{E} = \frac{1}{g} (\bs{s}_x \cdot \bs{s}_{x+\mu}  - 1) 
\end{equation*}
and expand it in the coupling $\hat{g} = g N$
\begin{equation}
\hat{E} = \frac{1}{g} (\bs{s}_x \cdot \bs{s}_{x+\mu}  - 1) = \hat{E}^{(0)} +\, \hat{g}\, \hat{E}^{(1)} +\, \hat{g}^{2} \hat{E}^{(2)} +\, \ldots 
\label{eq:hatE}
\end{equation}
The tree level result is now the zero order in the expansion, while the loop contributions exhibit a constant term plus $1/N$ corrections (the least significant power being $1/N^l$ al loop $l$; see Eq. \eqref{eq:know_res}); the absolute value of the $n$ order coefficient in turn no longer displays the $N^n$ growth. Fig. \ref{fig:MAINmessageNEW} is the new version of Fig. \ref{fig:MAINmessage}.

\begin{figure}[ht]
  \centering
  {\includegraphics[width=0.999\linewidth]{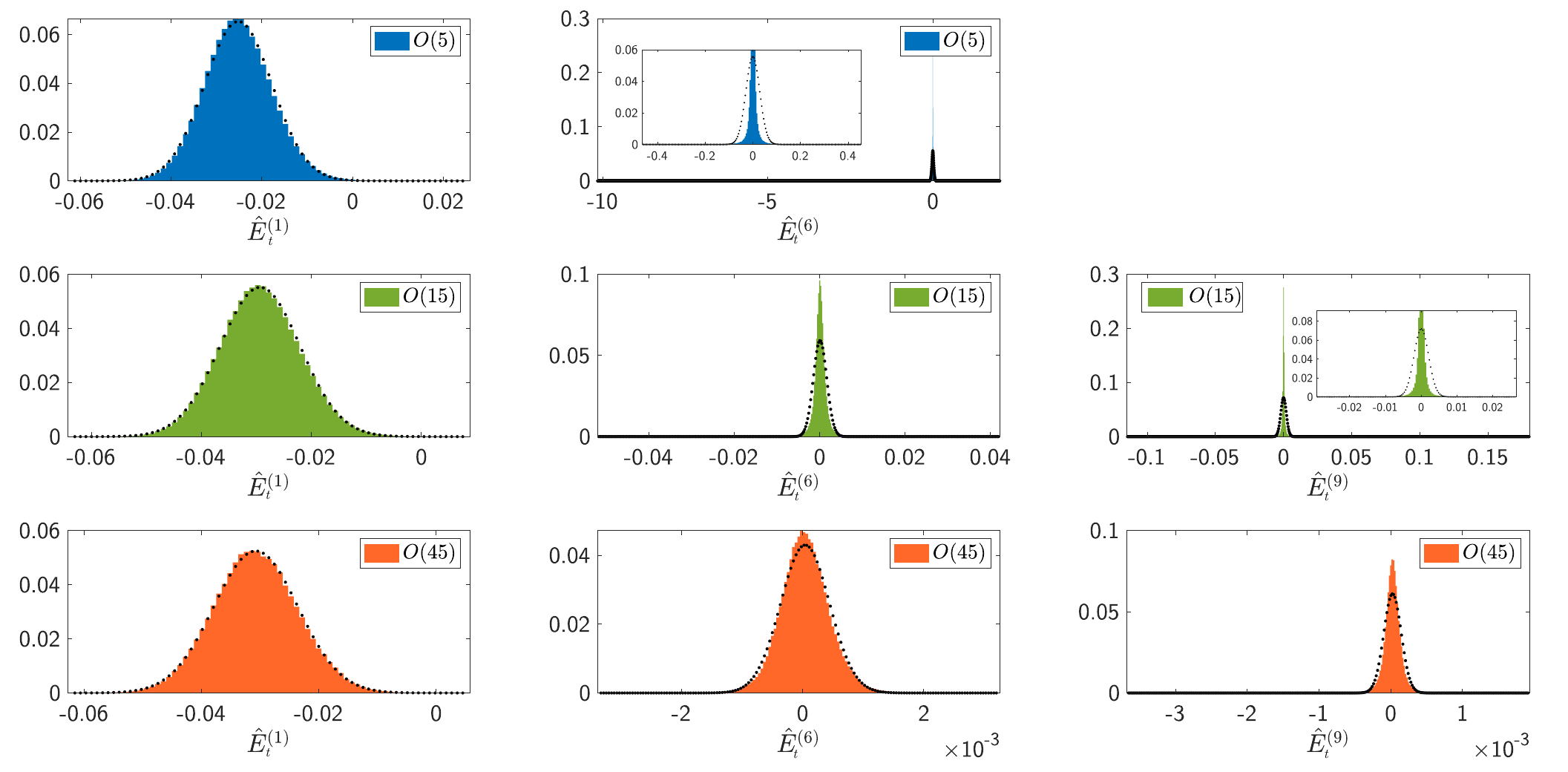}}
  \caption{The distributions of a few $\hat{E}_t^{(n)}$ for different values of $n$ and $N$: the choice of the $N$ values and color code remain the same (blue color refers to $O(5)$, green to $O(15)$, orange to $O(45)$). The distributions are compared to gaussian ones with the same mean values and standard deviations. For $N=5$ and $N=15$ the highest values of $n$ displayed are the largest for which the determination of mean values and standard deviations is reliable (see the discussion in section \ref{sec:fluctuations}).}
  \label{fig:histORDERS}
\end{figure}

In NSPT the coefficients $\hat{E}^{(n)}$ are evaluated by computing the average over an asymptotically large stochastic time interval of corresponding stochastic processes $\hat{E}_t^{(n)}$\footnote{More on this in section \ref{sec:StochProcAnalysis}}, the errors on the coefficients being controlled by the standard deviations of the processes and relevant autocorrelations. To compare fluctuations of the stochastic processes $\hat{E}_t^{(n)}$ at different $N$ and perturbative orders $n$ we now plot their distributions (which are somehow easier to inspect than those for the $E_t^{(n)}$, given the magnitude of the $\hat{E}^{(n)}$). Not only we can directly inspect the huge fluctuations showing up as long tails, but we can also inspect how these distributions deviate from the gaussian distribution; this is in the same spirit of \cite{GonzlezArroyo2022} (by which part of the analysis which will follows is to a certain extent inspired). Results are presented in Fig. \ref{fig:histORDERS} ($N$ fixed on each row; $n$ fixed on each column).\\

In Fig. \ref{fig:histORDERS} the distribution of each $\hat{E}_t^{(n)}$ is compared to the gaussian having the same mean value and standard deviation. First column refers to $n=1$: for all values of $N$ (all rows, $N=5,15,45$ respectively) the two distributions are quite well compatible. The second column refers to $n=6$: this is the largest perturbative order for which mean value and standard deviation are reliably computed for $N=5$ (see discussion in section \ref{sec:fluctuations}) and that's why there is nothing on the first row in the third column. 
Third column refers to $n=9$, which is in turn the largest perturbative order for which mean value and standard deviation are reliably computed for $N=15$. All in all, we can inspect on each row (fixed $N$) larger deviations from gaussianity (with larger tails in distributions) as perturbative order $n$ increases and on each column (fixed perturbative order $n$) smaller deviations from gaussianity as $N$ increases. For $N=5$ at $n=6$ the computation is still reliable, but the distribution of $\hat{E}_t^{(6)}$ is far far away from a gaussian distribution. For $N=15$ the distribution of $\hat{E}_t^{(6)}$ is already quite far away from a gaussian distribution; the deviation from gaussianity is huge at $n=9$ (and for larger perturbative orders the computation is no longer reliable). In the last row we can inspect what's going on for $N=45$; also in this case deviations from the gaussian distribution are larger for larger values of $n$, but all in all even at $k=9$ the distribution is much closer to the gaussian than for the other values of $N$. This tendency to gaussianity for larger $N$ is well compatible with general expectations, as we are inspecting systems with larger and larger number of degrees of freedom. \\

One would be interested in understanding whether the variances of the $\hat{E}_t^{(n)}$ stochastic processes can be related to field theoretic results, in particular to the (field theoretic) variance of the observable $\hat{E}$. In section \ref{sec:StochProcAnalysis} we will see that the answer is yes, but only to a given (particular) extent. Before we get there, we now try to assess how our perturbative results relate to (standard) large $N$ analysis.

\subsection{Deeper into large \texorpdfstring{$N$}{} analysis}

\begin{figure}[ht]
  \centering
  \includegraphics[width=1.1\linewidth]{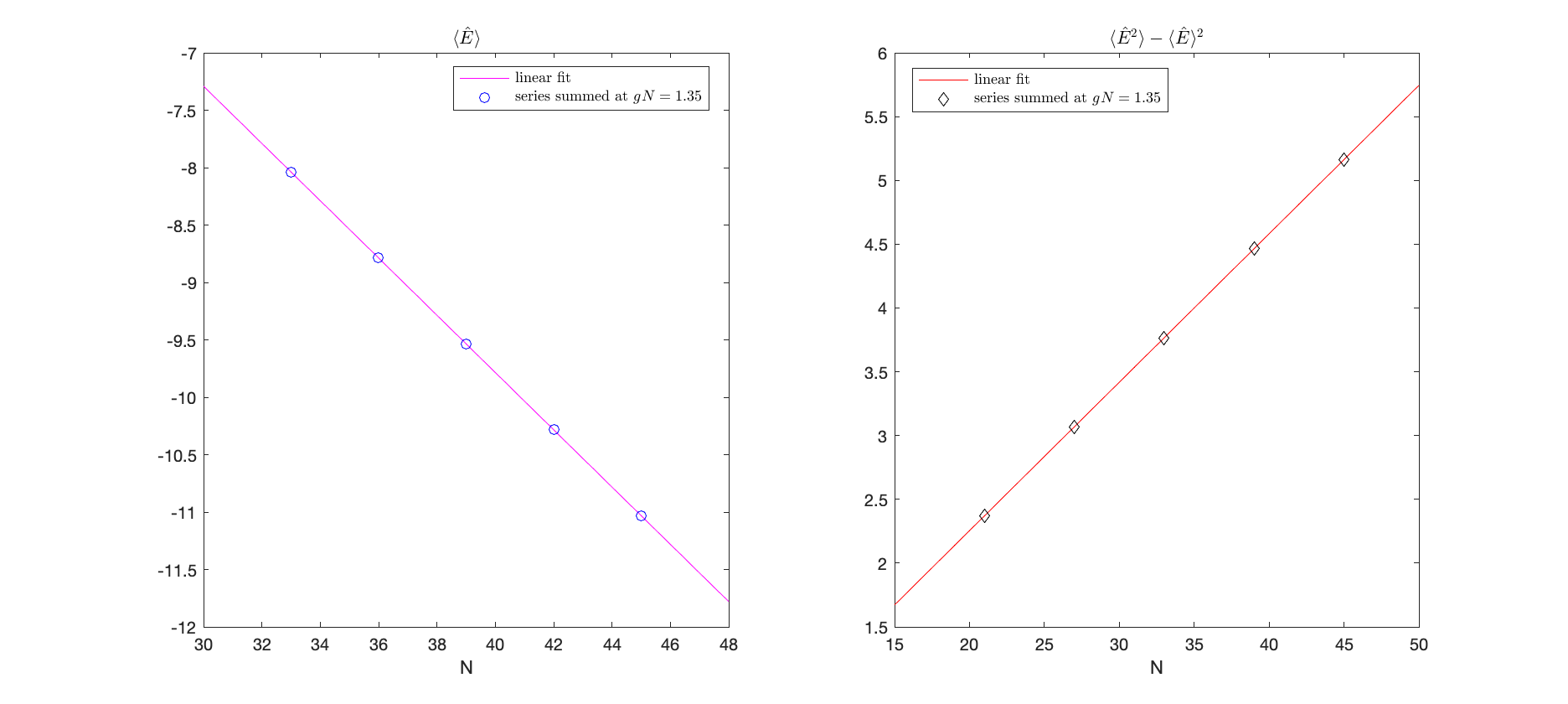}
  \caption{The observable $\langle \hat{E} \rangle$ (left panel) and the variance $\langle \hat{E}^2 \rangle - \langle \hat{E} \rangle^2$ (right panel) as obtained by summing our perturbative series for several values of $N$ at $\hat{g}=gN=1.35$ (see text for details on the summing prescription). They display agreement with the expected (linear) $N$ behavior.}
  \label{fig:SummingE}
\end{figure}

$O(N)$ are prototype models one can study in the large $N$ limit (see \cite{Makeenko:1999hq}, which we will to a large extent follow). We start from the continuum version of the model
\begin{equation*}
    S = \frac{1}{2g} \int dx \,(\partial_{\mu}\bs{s}(x))^2 \;\;\;\;\;\;\;\;\; \bs{s}(x)\cdot\bs{s}(x) = 1 \, \footnote{The reader will have already noticed that from time to time we adhere to the sloppy notation  $\bs{s}(x)\cdot\bs{s}(x) = \bs{s}(x)^2$.}
\end{equation*}
By a redefinition of the $\bs{s}(x)$ field, the coupling $g$ can now disappear from the action and reappear in the constraint 
$\delta(\bs{s}\cdot\bs{s} - 1/g)$, which is the most direct way of seeing that $g \sim 1/N$ ($\bs{s}$ has $N$ components). In writing the partition function, the constraint can in turn be enforced via an integral representation of the $\delta$-function  which results into the introduction of a(n auxiliary) field $u(x)$
\begin{equation}
\label{eq:newZ}
    Z \propto \int_{\uparrow} Du(x) \, \int D\bs{s}(x) \, \mbox{e}^{-\frac{1}{2} \int d^2x' \, \left[ (\partial_{\mu}\bs{s}(x'))^2 - u(x') \, (\bs{s}(x')^2 - \frac{1}{g}) \right] }  
\end{equation}
where $\int_{\uparrow} Du(x)$ means that we are integrating on a contour parallel to the imaginary axis. By performing the gaussian integration on the $\bs{s}$ fields we end up with
\begin{equation}
\label{eq:uField}
    Z \propto \int_{\uparrow} Du(x) \, \mbox{e}^{-\frac{N}{2} \, \mbox{Tr ln} \, \left( -\partial_{\mu}^2 + u(x) \right) - \frac{1}{2g} \int d^2x' \, u(x') }  
\end{equation}
From Eq. \eqref{eq:uField}) it follows that
\begin{itemize}
    \item The description of the theory in terms of the field $u(x)$ makes it possible to solve it in the large $N$ limit via saddle point approximation (see the $N$ in front of the effective action coming from the Jacobian of the transformation which diagonalized the original quadratic form in \eqref{eq:newZ}).
    \item The saddle point equation (in momentum space) reads
    \begin{equation}
        \frac{1}{g} = N \,\int^{\Lambda} \frac{d^2k}{2\pi^2} \frac{1}{k^2 + m_R^2}
    \end{equation}
    where one can read the translationally invariant solution $u_{SP}(x) = m_R^2 = \Lambda^2 \mbox{e}^{- \frac{4 \pi}{N g}}$, which is the mass generated via dimensional transmutation.
    \item In terms of the $u(x)$ field, in the large $N$ approximation, to the leading order in $1/N$ we get the factorization
    \begin{align}
    \begin{split}
    \label{eq:factorization}
        \langle u(x_1) u(x_2) \ldots u(x_n) \rangle & = \langle u(x_1) \rangle \langle u(x_2) \rangle \ldots \langle u(x_n) \rangle + \mathcal{O}(\frac{1}{N})\\
        & =  u_{SP}(x_1) \, u_{SP}(x_2) \ldots u_{SP}(x_n) + \mathcal{O}(\frac{1}{N})
    \end{split}
    \end{align}
    where we saw that $u_{SP}(x)$ is actually a (translationally invariant) constant.
\end{itemize}
We now see what all this translates into for our lattice $O(N)$ action, which can be rewritten
\begin{equation}
    S = - \frac{1}{g} \sum_{x,\mu} (\bs{s}_x \cdot \bs{s}_{x+\mu} - 1) = - \frac{1}{2g} \sum_{x,\mu} \bs{s}_x \cdot (\bs{s}_{x+\mu} + \bs{s}_{x-\mu} - 2 \bs{s}_{x} )
\end{equation}
Written in this way, the kernel defined by the action is symmetric and all the steps we took in the continuum case can be repeated in much the same way, ending up with
\begin{equation}
\label{eq:newZlat}
    Z \propto \int_{\uparrow} \prod_x du_x \, \mbox{e}^{-\frac{1}{2g} \sum_x u_x} \int \prod_{i,x} ds_{ix} \, \mbox{e}^{-\frac{1}{2} \sum_{i,j,x,y} \, s_{ix} \, \delta_{ij} K(u)_{xy} \,s_{jy}},  
\end{equation}
$K(u)$ being the lattice counterpart of the $(-\partial_{\mu}^2+u)$ continuum operator. To make contact with our computation, we first of all notice that our observable $\hat{E} = \frac{1}{g} \langle \bs{s}_x \cdot \bs{s}_{x+\mu} - 1 \rangle = \frac{1}{g} \langle \bs{s}_x \cdot \bs{s}_{x+\mu} - \bs{s}_x \cdot \bs{s}_x \rangle$ maps into $\langle \bs{s}_x \cdot \bs{s}_{x+\mu} - \bs{s}_x \cdot \bs{s}_x \rangle$ in the formulation of Eq. \eqref{eq:newZlat}. We then 
\begin{itemize}
    \item re-express $\langle \bs{s}_{x} \cdot \bs{s}_{x+\mu} - \bs{s}_x \cdot \bs{s}_x \rangle = \langle \sum_i s_{i x} s_{i x+\mu} - \sum_j s_{j x} s_{j x} \rangle$ in terms of the variables $z_{ix} \equiv \sum_y \sqrt{K}_{xy} s_{iy}$ (the ones introduced to diagonalize the quadratic form in the action),
    \item perform the gaussian integrals (which basically results in contractions à la Wick),
    \item and finally evaluate in saddle point approximation the $u$-integral we are left with, ending up with the result
\end{itemize}
\begin{equation}
\label{eq:LargeNres}
    \langle \sum_i s_{i x} s_{i x+\mu}  - \sum_j s_{j x} s_{j x} \rangle = N \, (K(u_{SP})^{-1}_{x \, x+\mu} - K(u_{SP})^{-1}_{x \, x})
\end{equation}
We stress that of course here we have been concerned with the leading order of a perturbative expansion in $\frac{1}{N}$, while we have previously presented our results for the perturbative expansion in $\hat{g}=gN$. Still, we have many orders in the latter, in particular for large enough values of $N$. Therefore, we can try to sum the series for large $N$ and look for the behavior dictated by Eq. \eqref{eq:LargeNres}. We adhere to a standard recipe: at a given value of $\hat{g}$, we look for the turning point in the sequence of the $\hat{E}_n \hat{g}^n$, {\em i.e.} we look for the largest $\Bar{n}$ for which $\hat{E}_{\Bar{n}} \, \hat{g}^{\Bar{n}} < \hat{E}_{\Bar{n}+1} \, \hat{g}^{\Bar{n}+1}$ and sum the series up to the $\Bar{n}$-th order. Fig. \ref{fig:SummingE} (left panel) shows that indeed we find the expected behavior.
While Fig. \ref{fig:SummingE} appears to display a remarkable agreement of the $\hat{E}$ observable (as obtained summing our perturbative expansions) with the expected $N$-behavior, this should not be taken as the end of the story. In particular, the figure has been obtained at a given value of the coupling $\hat{g}=gN$. The slope of the straight line should depend on the coupling via the dependence of the saddle point equation solution $m_R$ on $\hat{g}$, which in turn asks for a great control on the summation process. Summing our high orders perturbative series is a delicate issue, since they are supposed to be asymptotic series: all in all, one expects to deal with uncertainties. Here we are happy enough having inspected the linear behavior and we do not go deeper into these issues, since they will be the subject of another work \cite{ON_renormalons}. \\

Notice that one can also look for the large $N$ behavior of the variance of our observable $\hat{E}$. When factorization is in place,  for a given observable $O$, in the large $N$ limit, one could expect $\langle O^2 \rangle \sim \langle O \rangle^2$. This entails the vanishing of the field theoretic variance\footnote{We stress that this is the variance dictated by field theory, {\em i.e.} by the measure of the functional integral.} (again, in the large $N$ limit). Notice that this type of scenario is coded in Eq. \eqref{eq:factorization}, but we have already seen that in terms of the $s_{ix}$ field factorization takes place in the form dictated by contractions à la Wick. Performing the computation of $\langle (\sum_i s_{i x} s_{i x+\mu}  - \sum_j s_{j x} s_{j x})^2 \rangle - \langle \sum_i s_{i x} s_{i x+\mu}  - \sum_j s_{j x} s_{j x} \rangle^2$ along the same lines that took us to Eq. \eqref{eq:LargeNres} one finds that the $N^2$ terms cancel, and one is again left with a behavior that is linear in $N$. This is exactly what we can inspect in the right panel of Fig. \ref{fig:SummingE}, where at a given value of $\hat{g}$ we sum the perturbative series of NSPT perturbative computations of the variance of $\hat{E}$ at several values of $N$ (the sums have been obtained by the same procedure we described above). \\

\begin{figure}[ht]
  \centering
  {\includegraphics[width=0.6\linewidth]{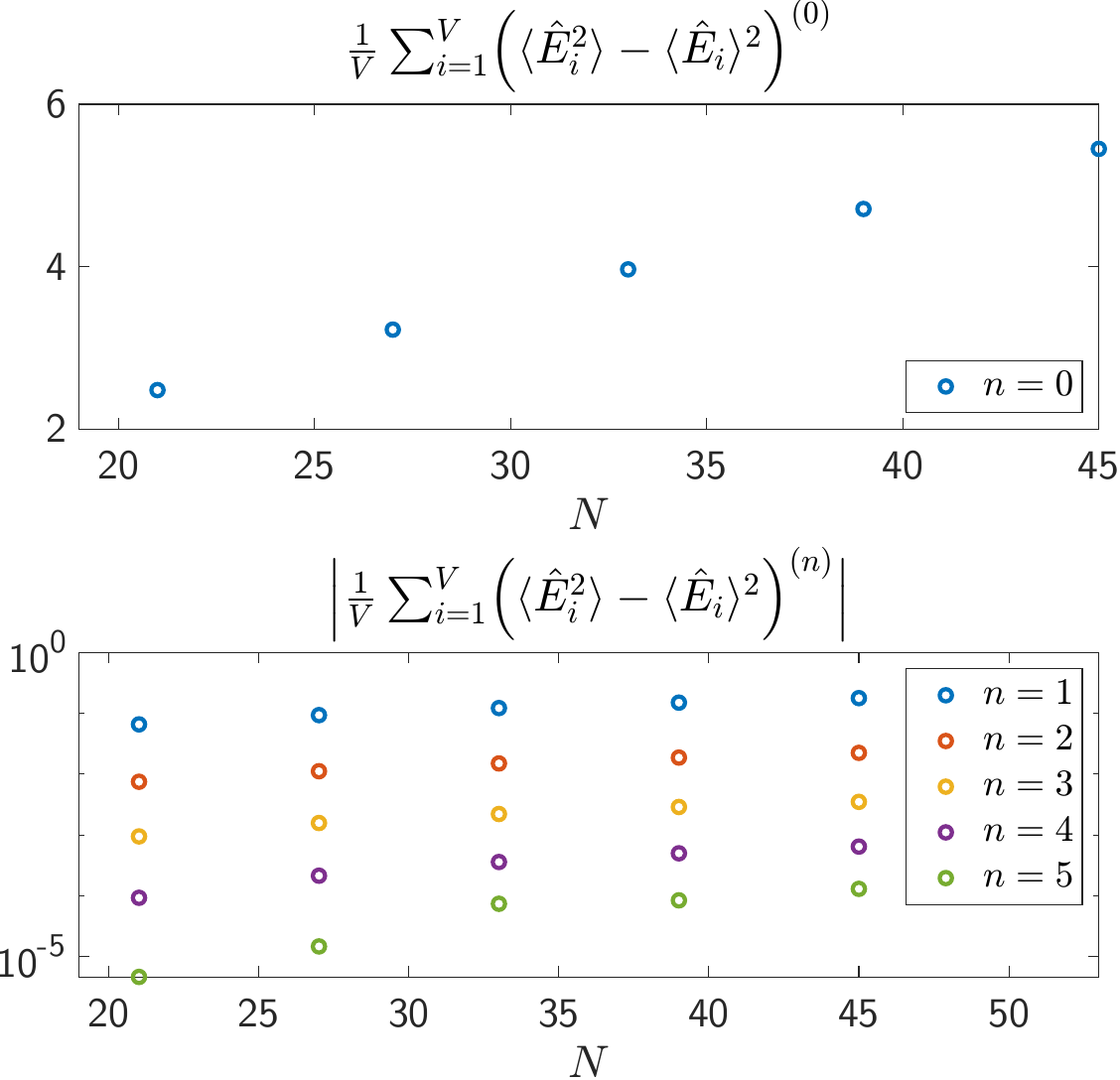}}
  \caption{Perturbative coefficients of the variance $(\langle \hat{E}^2 \rangle - \langle \hat{E} \rangle^2)^{(n)}$ for a number of $N$ values ($x$-axis): in the upper panel, the tree level contribution; in the lower panel, the absolute value of loop contributions, which are all negative, that is, they reduce the value of the variance, when you sum the series. The variance is computed at a generic point and averaged over the entire lattice.}
  \label{fig:varQFT}
\end{figure}

\begin{figure}[hbt]
  \centering
  {\includegraphics[width=0.6\linewidth]{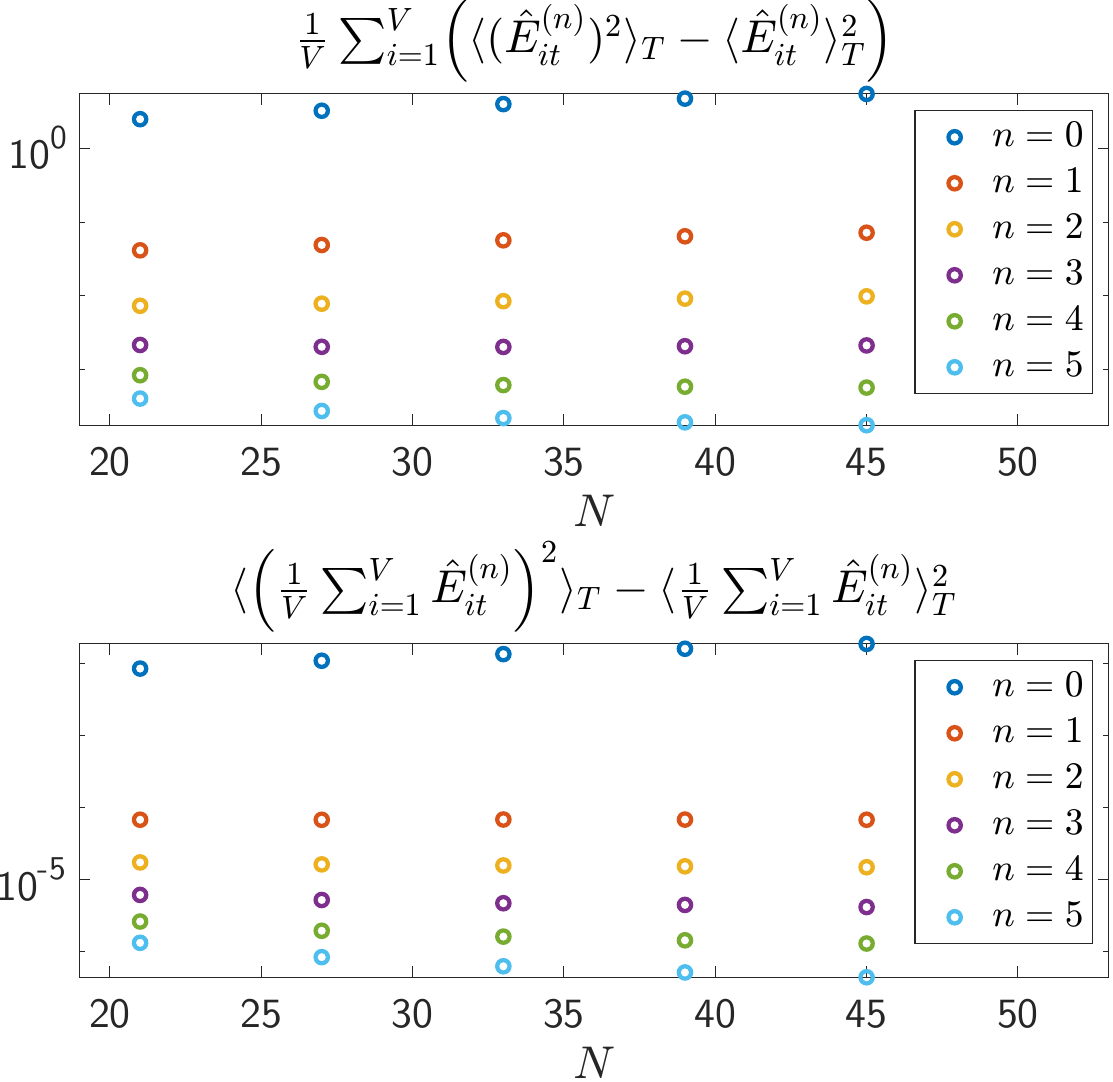}}
  \caption{Variances of a few NSPT processes $\hat{E}_t^{(n)}$. We display both the lattice average of the variance of processes computed at a generic point and the (smaller) variance of lattice averaged processes. In both cases, for $n>1$ the variances get smaller for larger $N$.}
  \label{fig:varNSPT}
\end{figure}

We now focus on what we are mainly interested in, {\em i.e.} the perturbative computation of the variance $D_{\hat{E}} = \langle \hat{E}^2 \rangle - \langle \hat{E} \rangle^2$. The results we got (from NSPT) are displayed in Fig. \ref{fig:varQFT}. Notice that loop corrections are all negative, that is,  by summing the series the variance gets reduced with respect to the tree-level value. 
We said that inspecting the variance of $\hat{E}$ is our main interest because we want to go back to our (original) question: are the variances of the stochastic processes $\hat{E}_t^{(n)}$ any related to the perturbative coefficients of the field theoretic variance that we have just (computed and) displayed? The variance of a few NSPT processes $\hat{E}_t^{(n)}$ are plotted in Fig. \ref{fig:varNSPT}\footnote{We plot variances of processes for both the lattice average of the observable computed at a generic point and the lattice averaged observable, the second being of course much smaller.}. We get yet another confirmation that as the order $n$ increases, the variances get smaller for larger $N$. 
Apparently there is not that much similar when we compare Fig. \ref{fig:varQFT} and Fig. \ref{fig:varNSPT}, so that one could think there is no point in the question we asked ourselves. This is not really true, and in the next section we will derive a general result.

\section{Field theoretic variance and the variance of the \texorpdfstring{$\hat{E}^{(n)}$}{}}
\label{sec:StochProcAnalysis}

We are going to discuss an important (general) property of the variance of NSPT processes. At each stochastic time $t$ an observable $O$ is associated to a sum of stochastic processes, one for each order we are taking into account\footnote{We use $\hat{g}$ as the coupling, since this the one at hand in this context; the discussion is of course a general one.}

\begin{equation*}
\label{eq:operator_t}
    O_t = O_t^{(0)} +\, \hat{g}\, O_t^{(1)} +\, \hat{g}^2 O_t^{(2)} +\, \ldots
\end{equation*}

To keep the notation light we are discarding any dependence on fields or (other) indexes or space-time coordinates, retaining only the dependence on the stochastic time, which we write as a subscript. For example, our observable $\hat{E}$ has a dependence on a space-time coordinate, which is typically averaged over (due to translational invariance). All the $O_t^{(j)}$ are constructed as composite operators, built out of the different perturbative orders of the field: for any theory, the zeroth order $O_t^{(0)}$ is only a function of the zeroth order field $\phi_t^{(0)}$, the first order $O_t^{(1)}$ is a function of the zeroth order field $\phi_t^{(0)}$ and the first order field $\phi_t^{(1)}$, and so on. In our case, fields have components ($s_i$, which translates into $s_i^{(j)}$), but once again the fact that those are there and are contracted into a scalar product is not manifest in Eq. \eqref{eq:operator_t}, in which we only stress that all the composite operators $O_t^{(j)}$ are computed at (the same) stochastic time $t$. The perturbative series for the observable $O$ is computed in NSPT as
\begin{align}
\begin{split}
\label{eq:TIMEaverages}
    \langle O \rangle  & =  c_0 +\, c_1\, \hat{g}\, +\, c_2\, \hat{g}^2 +\, \ldots \\
     & = \langle O_t^{(0)} \rangle_T +\, \hat{g}\, \langle O_t^{(1)} \rangle_T +\, \hat{g}^2\langle O_t^{(2)} \rangle_T +\, \ldots\\
\end{split}
\end{align}
where the notation $\langle \ldots \rangle_T$ has the meaning 
\begin{equation}
    \langle O_t^{(j)} \rangle_T \equiv \frac{1}{T} \sum_{t=1}^T O_t^{(j)}\footnote{We remind that for asymptotically large $T$ this equals $\lim_{t \rightarrow \infty} \langle O_t^{(j).} \rangle_{\eta}$}
\end{equation}
that is, coefficients of a perturbative expansion are computed in NSPT as time averages (over an asymptotically large time interval $T$) of the corresponding (composite) operators processes. The errors on the $c_j$ are computed by estimating the variance of the (stochastic time) averages $\langle O_t^{(j)} \rangle_T$ and the relevant autocorrelations. We will now relate the variances of the processes with the perturbative expansion of the field theoretic variance of $O$.\\
We compute the field theoretic variance in perturbation theory by evaluating $D_O = \langle O^2 \rangle - \langle O \rangle^2$ in NSPT. The composite operators contributing to $O^2$ are easily computed (in the order-by-order formalism which is the key point of NSPT)\footnote{For the sake of simplicity we take as an example a case like the one at hand in this work, where there is no subtlety due to non-commutative nature.}
\begin{equation}
    O^2_t = O_t * O_t = O_t^{(0)\,2} +\, \hat{g}\, (2\, O_t^{(0)} O_t^{(1)}) +\, 
    \hat{g}^2 (O_t^{(1)\, 2} + 2\, O_t^{(0)} O_t^{(2)}) +\, \ldots
\end{equation}
and thus
\begin{align}
\begin{split}
\label{eq:TIMEaverages_sq}
    \langle O^2 \rangle  & =  c_{2 0} +\, c_{2 1}\, \hat{g}\, +\, c_{2 2}\, \hat{g}^2 +\, \ldots \\
     & = \langle O_t^{(0)\,2} \rangle_T +\, \hat{g}\, \langle 2\, O_t^{(0)} O_t^{(1)} \rangle_T +\, 
     \hat{g}^2\langle O_t^{(1)\, 2} + 2\, O_t^{(0)} O_t^{(2)} \rangle_T +\, \ldots\\
\end{split}
\end{align}
The order by order evaluation of the variance is obtained by taking the previous result and subtracting the (perturbative expansion of) the square of $\langle O \rangle$, that is
\begin{equation}
    D_O = \langle O^2 \rangle - \langle O \rangle^2 = (c_{2 0} - c_0^2) \, +\, \hat{g}\, (c_{2 1} - c_0 c_1) \,+\,
    \hat{g}^2 (c_{2 2} - 2 c_0 c_2 - c_1^2) \,+\, \ldots
\end{equation}
Collecting everything together we find
\begin{align}
\begin{split}
\label{eq:CorrelVar}
    (\langle O^2 \rangle - \langle O \rangle^2)^{(0)}  & =  \langle O_t^{(0)\,2} \rangle_T - \langle O_t^{(0)} \rangle_T^2 \\
     (\langle O^2 \rangle - \langle O \rangle^2)^{(1)}  & = 2\, (\langle O_t^{(0)} O_t^{(1)} \rangle_T - 
     \langle O_t^{(0)} \rangle_T \langle O_t^{(1)} \rangle_T) \\
     (\langle O^2 \rangle - \langle O \rangle^2)^{(2)}  & = 2\, (\langle O_t^{(0)} O_t^{(2)} \rangle_T - 
     \langle O_t^{(0)} \rangle_T \langle O_t^{(2)} \rangle_T) \,+\, 
     \langle O_t^{(1)\,2} \rangle_T - \langle O_t^{(1)} \rangle_T^2 \\
     \ldots & \\
\end{split}
\end{align}
In inspecting the tower of equalities we should keep in mind that at each perturbative order on the left hand side we read quantities that are fixed by field theory (and thus have nothing to do with how we compute them), while on the right hand side we read quantities that characterise stochastic properties of our NSPT processes. \\
We can see that the zeroth order of the field theoretic variance equals the stochastic variance of $O_t^{(0)}$. The first order of the field theoretic variance in turn equals twice the covariance of the NSPT processes $O_t^{(0)}$ and $O_t^{(1)}$. We can then see that the second order of the field theoretic variance equals the sum of twice the covariance of the NSPT processes $O_t^{(0)}$ and $O_t^{(2)}$ plus the variance of the NSPT process $O_t^{(1)}$. From these first three orders we can inspect all the general properties of this tower of equalities. In particular, at any even order $n=2k$ the nth order of the field theoretic variance equals the sum of the variance of the NSPT process $O_t^{(n/2)}$ coefficient plus twice the sum of all the covariances of the NSPT processes $O_t^{(l)}$ and $O_t^{(m)}$ with $l+m=n$. 
Notice that the variances appearing on the right hand side are just those we are interested in (in the end, we want to compute the error associated to the computation of $c_j$ as stochastic time average of the process $O_t^{(j)}$). Since these variances are the only positive definite quantities around, at each order twice the sum of the relevant covariances are less or equal than the field theoretic variance at that given order (in particular, an equality holds at odd orders). In our case ({\em i.e.} taking $\hat{E}$ as the observable $O$) the perturbative coefficients of the field theoretic variance are all negative, and indeed we find that all those relevant sums of covariances are negative as well. Notice that all this is interesting also with respect to the (different) equivalent formulations \cite{DallaBrida2017, DallaBrida20172} of NSPT, {\em i.e.} those based on stochastic equations different from Langevin. We stress once again that what is universal are the field theoretic results on the left hand side: given an observable, the different perturbative orders of its variance are given (although, of course, difficult to compute at high orders if one does not rely on NSPT). On the other hand, we can conclude that
\begin{itemize}
    \item not only the stochastic variance of the (leading) $O_t^{(0)}$ is universal, but also the covariance of the processes $O_t^{(0)}$ and $O_t^{(1)}$: these must be the same whatever variant of NSPT one implements;
    \item on the other side, only the combination of the variance of the $l$th order stochastic process and of certain covariances of stochastic processes (at given orders) is universal at any order $l \geq 1$. 
\end{itemize}
Although we expect that the {\em smaller variances for a larger number of internal degrees of freedom} statement will hold true for all the variants of NSPT, the latter point implies that (maybe even quite) different numbers for NSPT variances show up at different orders for different NSPT variants, but always subject to the constraints coming from covariances. This is the subject of another work of ours \cite{NSPT_variants_variances}.
\color{black}
\section{Conclusions and prospects}

We made use of NSPT to compute the energy of $O(N)$ Non Linear Sigma Models, for several increasing values of $N$. Our first goal was to establish that indeed for relatively {\em small} number of {\em internal} degrees of freedom we would find increasing large fluctuations for NSPT computations at increasing perturbative orders $n$, actually so large that at some point the signal to noise ratio is not under control. At the same time, we were able to verify a simple conjecture: NSPT high order computations are expected to be safe when we deal with a {\em large} number of {\em internal} degrees of freedom. Our main conclusion is that high orders fluctuations are tamed for {\em large} $N$: basically, the larger is $N$, the larger is the perturbative order $n$ that we can safely compute. Our understanding for this is an admittedly simple one, which could be tested by comparing the distributions of the stochastic processes at different perturbative orders for different $O(N)$: we found a tendency to gaussianity for larger $N$. This is well compatible with general expectations, as we are inspecting systems with larger and larger number of degrees of freedom.
We found that {\em a posteriori} it was possible to state how large $N$ must be to safely compute at perturbative order $n$. Different indicators for the selection of such thresholds result in the same conclusions. \\

Not surprisingly, the study of the distributions of the processes was much easier by switching to a coupling à la 't Hooft. Having computed several orders in the perturbative expansions in the $\hat{g}=gN$ coupling, it was possible to sum our perturbative series and compare our results with leading order results of large $N$ analysis \cite{Makeenko:1999hq}. It would be interesting to tackle comparisons also with other results obtained by standard large $N$ continuum techniques \cite{Brezin1976, Brezin1976b, Bardeen1976}. \\

We also established a general result, which is valid for any variant of NSPT. For any given observable $O$, there are relationships between the various perturbative orders of the field theoretic variance of $O$ and variances and covariances of the stochastic processes NSPT deals with. In particular, the (even) $2k$ orders in the expansion of the variance of $O$ equals the sum of the variance 
of the $k$ order stochastic process $O_t^{(k)}$ plus the sum of a given number of covariances of stochastic processes $O_t^{(m)}$ and $O_t^{(l)}$, with $l+m=2k$. This motivated us to tackle a comparative study of different variants of NSPT \cite{NSPT_variants_variances}
.\\

Once we have found that NSPT computations can be safely performed for large enough $N$, we have the chance to inspect asymptotics. This is due if we want to better control the procedure of summing our (asymptotic) series, but this is not the only issue to address. In particular, we can study the expected IR-renormalon behavior. This is what we are doing in \cite{ON_renormalons} (and the study of renormalones is performed fully accounting for finite volumes effects). \\

The present work has been in part motivated by problems encountered in NSPT computations around non trivial vacua (double well potential in Quantum Mechanics \cite{Baglioni2023}). The idea is now to go for non-trivial vacua NSPT computations in $CP(N-1)$ models, which were among the very first related to resurgence scenarios \cite{Dunne:2012ae}.

\section *{Acknowledgments}
We thank Petros Dimopoulos for very interesting discussions. This work was supported by the European Union Horizon 2020 research and innovation programme under the Marie Sk\l odowska-Curie grant agreement No 813942 (EuroPLEx) and by the INFN under the research project (\textit{iniziativa specifica}) QCDLAT. This research benefits from the HPC (High Performance Computing) facility of the University of Parma, Italy.
\color{black}

\bibliographystyle{JHEP}
\bibliography{references.bib}

\newpage

\begin{appendices}

\section{Zero-mode regularization}
\label{app:zero_mode}

\begin{figure}[t]
\centering
  {\includegraphics[width=1.\linewidth]{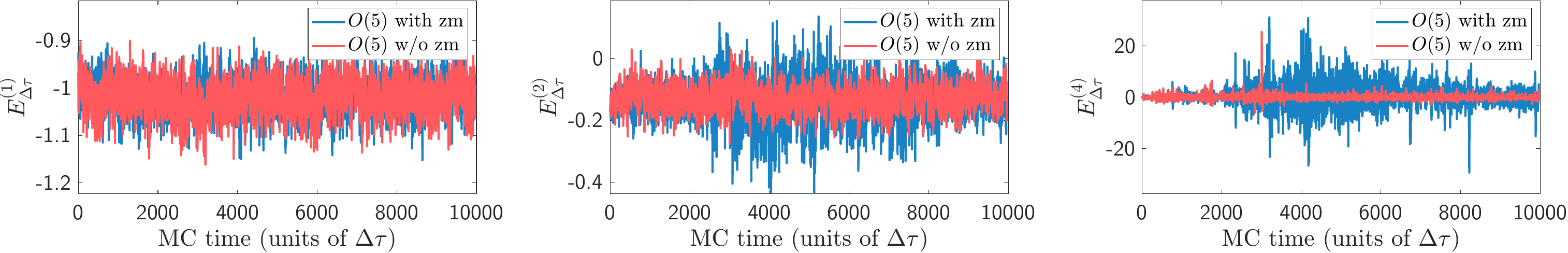}}
  \caption{NSPT time series for the energy of $O(5)$ NLSM model ($20\times 20$ lattice). Different panels represent increasing loop orders, with (blue lines) and without (red lines) subtraction of zero-mode. The expected on average cancelation of divergence is taking place, but (as expected) one pays a huge noise price.}
  \label{fig:zeromode}
\end{figure}

Given the action in Eq. \eqref{eq:action}, it is straightforward to verify that the basic building block of the diagrammatic perturbation theory, the propagator
\begin{equation}
\label{eq:prop}
    \braket{\bs{\pi}_x \cdot \bs{\pi}_y} = \sum_{y,k} \delta_{yk} \braket{\pi^y_x \pi^k_y}
\end{equation}
is ill-defined, having a zero-mode. An interesting feature is that if we consider only $O(N)$ invariant quantities, like the complete propagators
\begin{equation}
\label{eq:complete_prop}
    \braket{\bs{s}_x \cdot \bs{s}_y} = g \braket{\bs{\pi}_x \cdot \bs{\pi}_y} + \braket{\sigma_x \sigma_y}
\end{equation}
these are well-defined quantities because the last term in Eq. \eqref{eq:complete_prop} cancel order-by-order the zero-mode contribution \cite{Elitzur1983}. From the NSPT point of view, this cancellation is expected to take place on average, which means plagued by statistical noise. One possibility is to introduce an infrared regulator $\lambda$ in the same spirit of Eq. \eqref{eq:PF_PT} and to remove it via a $\lambda \rightarrow 0$ extrapolation. This results in additional computational load. Moreover, in our experience, to achieve good numerical signals one has to live with quite large values of $\lambda$. 
A simpler and more effective choice is to remove the zero mode order-by-order, subtracting this contribution directly from the field (a recipe also used for LGT \cite{DiRenzo2004}). This procedure is very cheap and in principle correct in the thermodynamic limit, while at finite volume it introduces extra finite size effects. As seen, these are nevertheless tested to be small (almost negligible up to the fourth loop order). Subtraction of zero mode and on average subtraction are compared in Fig. \ref{fig:zeromode} for the $O(5)$ model.

\section{\texorpdfstring{$\Delta\tau\to 0$}{} extrapolations}
\label{app:tau_estrap}

\begin{figure}[htb]
\centering
  {\includegraphics[width=1.\linewidth]{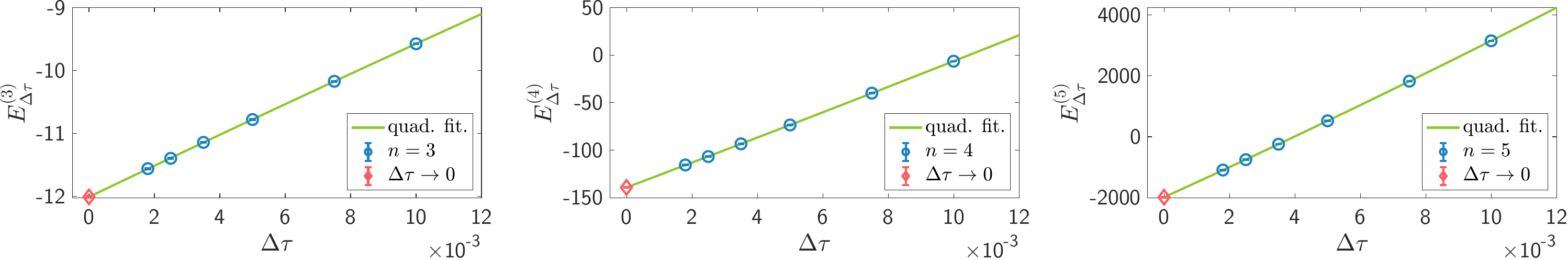}}
  \caption{Example of $\Delta\tau\to 0$ extrapolation in the $O(15)$ model for perturbative order $n=3,4,5$ (on a $20\times 20$ lattice); we obtain $\chi^2_{red.} \approx 1.33$. Blue markers represent measurements at different stochastic time step (notice that errors are too small to be seen), green lines are the quadratic fit estimated by means of Eq. \ref{eq:chi_sq} and red diamonds are the extrapolated quantities.}
  \label{fig:dt_estrap}
\end{figure}

\begin{figure}[hb]
\centering
  {\includegraphics[width=0.85\linewidth]{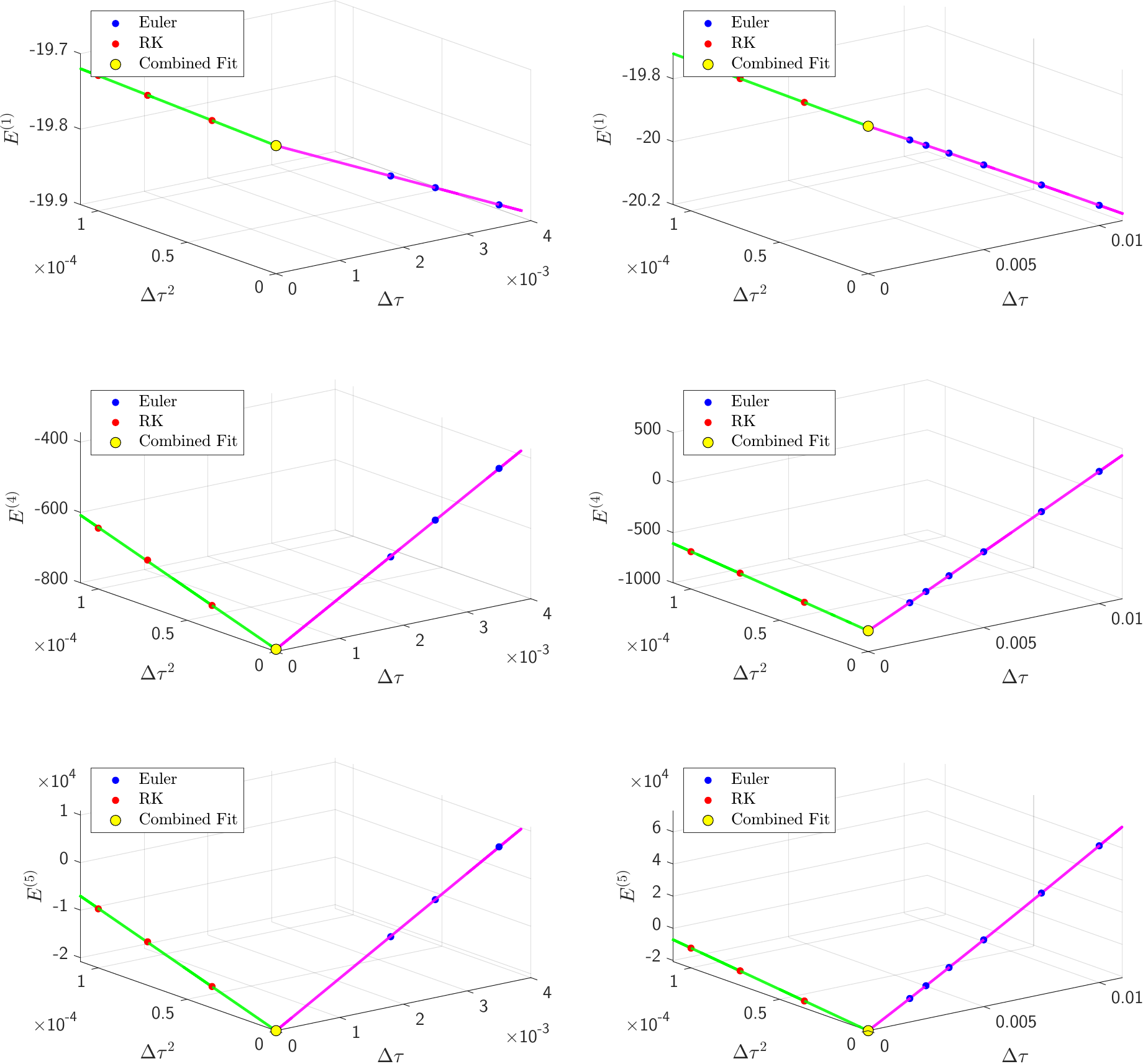}}
  \caption{{\em Two schemes} extrapolation procedure for the $O(80)$ NLSM. Three perturbative orders are displayed on the three rows. Both Euler (blue markers) and Runge Kutta (red markers) integration schemes were used. In the left column, the fitting procedure entails a linear extrapolation in $\Delta\tau$ for Euler (magenta solid line) and a linear extrapolation in $\Delta\tau^2$ for Runge-Kutta (green solid line). In the right column (same color code), a cubic extrapolations in $\Delta\tau$ for the Euler data is performed (taking into account more values of $\Delta\tau$). The values we are interested in are those of the intercepts (yellow markers), resulting from the combined fit.}
  \label{fig:dt_estrap2}
\end{figure}

Coefficients of different perturbative orders in an NSPT computation at a given $\Delta\tau$ are correlated, while at different $\Delta\tau$ values of course they are not. In force of that, the extrapolation to vanishing stochastic time step (see Fig.~\ref{fig:dt_estrap}) has been performed minimizing the generalised $\chi^2$ function \cite{DelDebbio2018}:
\begin{equation}
\label{eq:chi_sq}
    \chi^2 = \sum_{n,m,\Delta\tau}(E^{(n)}_{\Delta\tau} - \alpha_n \Delta\tau - \beta_n \Delta\tau^2 - E^{(n)}) \Sigma^{-1}_{\Delta\tau}(n,m) (E^{(m)}_{\Delta\tau} - \alpha_m \Delta\tau - \beta_m \Delta\tau^2 - E^{(m)})
\end{equation}
where it can be seen that both linear and quadratic effects in the time step can be (and have been) considered in the extrapolation.
The indexes run on perturbative orders $n,m \in [0,1,\dots,n_{max}]$, being $n_{max}$ the maximum perturbative order included in the fit. $\Sigma^{-1}_{\Delta\tau}(n,m)$ is the inverse of the covariance matrix at the same $\Delta\tau$ encoding the correlation between different perturbative orders $n$ and $m$. $E^{(n)}$ are the extrapolated quantities we are interested in. The block diagonal $\Sigma^{-1}_{\Delta\tau}(n,m)$ matrix encodes also the cross- and auto-correlations (see App. \ref{app:autocorr} for a detailed discussion). In addition the Gaussian Sampling method was used for error propagation. \\
The fact that we have to live with an extrapolation in the stochastic time has been quite debated. All in all, we are here making use of the simplest recipe, namely the Euler scheme. Other schemes are viable, which we also make use of (see the following). A key point is that we are here interested in high orders, so we want to make sure that extrapolation is safe, taking every perturbative order into account. One temptation to stay away from is that of thinking that for some higher order scheme a single value of the time step can be relied on. There is no way of getting a value of the time step which is {\em small enough} to ensure a precision which we cannot distinguish from statistical errors: higher orders can always hold surprises (and in general they do, at least because orders of magnitude can be very different at different orders). As an extra comment, notice that a stochastic time etrapolation turning out to be viable is yet another indication that a result at a given order is safe (from the point of view of huge fluctuations). In other words, stochastic time extrapolation does result in computational efforts, but it can make us more confident we can trust means and errors we compute at fixed values of $\Delta\tau$. This is in the end the main thing we are interested in here. \\
There are cases in which we actually do even more, that is we even make use of more than one numerical scheme. In Fig,~\ref{fig:dt_estrap2}
we show an example of a {\em two schemes} extrapolation procedure, which is performed to gain extra confidence in our results (this refers to computations for $O(80)$ on a $32 \times 32$ lattice, which are relevant for \cite{ON_renormalons}).

\section{Auto- and cross-correlations}
\label{app:autocorr}

\begin{figure}[ht]
\centering
  {\includegraphics[width=1.\linewidth]{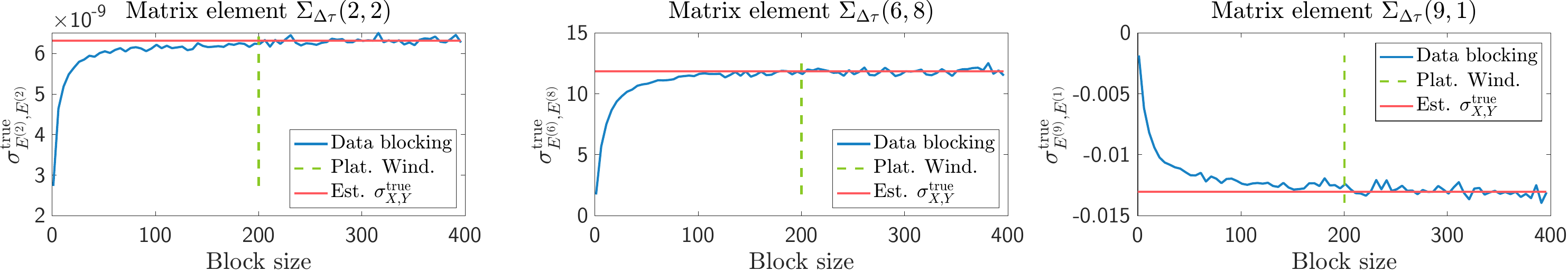}}
  \caption{Estimation of the matrix elements $\Sigma_{\Delta\tau}(n,m)$ for the $O(15)$ model at finite time step $\Delta\tau = 0.005$. The plots show the covariance values for each block size in blue. The true covariance was computed in the (plateau) region to the right of the green dashed line. The estimated covariance is the red horizontal line. Each panel displays the loop order pair $(n,m)$. All block sizes are multiple of 100.}
  \label{fig:blocking}
\end{figure}

The matrix $\Sigma^{-1}_{\Delta\tau}(n,m)$ was computed taking into account the auto- and cross-correlations between data evaluated at the same time step. To account for this, the blocking method was used \cite{Gattringer2010}. The different perturbative time series were organized into data blocks, averages were calculated on each block, and then the covariance matrix $\Sigma_{\Delta\tau}(n,m)$ was estimated. This procedure was repeated for increasing block sizes, until variances and covariances of the blocks reached a plateau. Once on the plateau, the block sizes were further increased in fixed steps, up to twice the size at which the plateau was (clearly) reached. The stability of the plateau could thus be verified, and the value of the true covariance $\sigma_{X,Y}^{\mathrm{true}}$ (or the variance $\sigma_{X,X}^{\mathrm{true}}$) was estimated as the average of the points in the plateau region. Estimations of the matrix elements $(2,2)$, $(6,8)$, $(9,1)$ in the $O(15)$ case are displayed in Fig. \ref{fig:blocking}. Additionally, from the value of the variance $\sigma_{X,X}^{\mathrm{true}}$ and the trivial variance $\sigma_{X}^2$ it is possible to derive the value of the integrated autocorrelation time $\tau_{int}$ \cite{Wolff2004}. There are various packages that offer an estimate of $\tau_{int}$ through the Gamma Functions method \cite{Ramos2019, Joswig2023} so that some results were double checked, finding full agreement.

\end{appendices}

\end{document}